\documentclass[
twocolumn,
aps,
prb,
reprint,
superscriptaddress,
amsmath,amssymb,
]{revtex4-1}
\usepackage[pdftex]{graphicx, color}

\usepackage[pdftex,
            colorlinks=true,
            citecolor=blue,
            linkcolor=blue]{hyperref}

\usepackage{braket,ulem}
\usepackage{times,multirow,amsfonts,bm,xspace,pifont}
\usepackage{soul}
\begin{document}
\newcommand{\rr}{{\bm r}}
\newcommand{\q}{{\bm q}}
\renewcommand{\k}{{\bm k}}
\newcommand*\wien    {\textsc{wien}2k\xspace}
\newcommand*\textred[1]{\textcolor{red}{#1}}
\newcommand*\textblue[1]{\textcolor{blue}{#1}}
\newcommand{\ki}[1]{{\color{red}\st{#1}}}
\newcommand{\sgn}{\mathrm{sgn}\,}
\newcommand{\tr}{\mathrm{tr}\,}
\newcommand{\Tr}{\mathrm{Tr}\,}
\newcommand{\GL}{{\mathrm{GL}}}
\newcommand{\talpha}{{\tilde{\alpha}}}
\newcommand{\tbeta}{{\tilde{\beta}}}
\newcommand{\mathN}{{\mathcal{N}}}
\newcommand{\mathQ}{{\mathcal{Q}}}
\newcommand{\bv}{{\bar{v}}}
\newcommand{\bj}{{\bar{j}}}
\newcommand{\zero}{{(0)}}
\newcommand{\one}{{(1)}}
\newcommand{\two}{{(2)}}
\newcommand{\three}{{(3)}}
\newcommand{\four}{{(4)}}

\newcommand{\YY}[1]{\textcolor{blue}{#1}}
\newcommand{\AD}[1]{\textcolor{red}{#1}}
\newcommand{\ADS}[1]{\textcolor{red}{\sout{#1}}}

\graphicspath{{./fig_submit/}}

\title{Superconducting diode effect and nonreciprocal transition lines
}

\author{Akito Daido}
\affiliation{Department of Physics, Graduate School of Science, Kyoto University, Kyoto 606-8502, Japan}
\email[]{daido@scphys.kyoto-u.ac.jp}
\author{Youichi Yanase}
\affiliation{Department of Physics, Graduate School of Science, Kyoto University, Kyoto 606-8502, Japan}
\date{\today}

\begin{abstract}
Nonreciprocity in superconductors is attracting much interest owing to its fundamental importance as well as the potential applicability to engineering.
In this paper, we generalize the previous theories of the intrinsic superconducting diode effect (SDE) and microscopically elucidate its relationship with the nonreciprocity of the transition lines under supercurrent.
We derive a general formula for the intrinsic SDE by using the phenomenological Ginzburg-Landau theory and thereby show that the SDE is determined by the relative angle between the magnetic field and an effective anti-symmetric spin-orbit coupling defined from the Ginzburg-Landau coefficients.
The obtained formula offers a convenient criterion to obtain a finite SDE.
We also study the SDE and the nonreciprocal phase transitions of the $s$-wave and $d$-wave superconductors by using the mean-field theory.
It is established that the sign reversal of the SDE accompanied by the crossover of the helical superconductivity is a general feature irrespective of the system details.
We study the phase transition lines in the temperature-magnetic-field phase diagram under the supercurrent, and
clarify that the sign reversal of the SDE generally accompanies the crossings of the transition lines under positive and negative current directions.
Furthermore, the superconducting phases under the supercurrent even become re-entrant under moderate strength of the electric current, implying the current-induced first-order phase transitions.
Our findings establish the electric current as the control parameter and the powerful probe to study the superconducting properties related to the finite-momentum Cooper pairs.
\end{abstract}

\maketitle

\section{Introduction}
Nonreciprocal phenomena in noncentrosymmetric materials are attracting much attention these days~\cite{Tokura2018-nb,Ideue2021-es}.
Nonreciprocity refers to the inequivalence of the left-mover and the right-mover: For instance, the nonreciprocity of the electric resistance is called the magnetochiral anisotropy, and has been observed in various materials~\cite{Rikken2001-il,Krstic2002-vo,Pop2014-wn,Rikken2005-ew,Ideue2017-vp,Wakatsuki2018-ll,Hoshino2018-sa,Wakatsuki2017-dp,Qin2017-vd,Yasuda2019-jw,Itahashi2020-ef}.
The experimental platform of the magnetochiral anisotropy includes superconductors near the transition temperature~\cite{Wakatsuki2017-dp,Qin2017-vd,Yasuda2019-jw,Itahashi2020-ef,Zhang2020-al}.
It has been pointed out that 
the spin-singlet and -triplet mixing of Cooper pairs can be detected~\cite{Wakatsuki2018-ll}, which is generally difficult to be identified.
Observation of nonreciprocal properties in materials, including nonlinear optical responses of superconductors~\cite{Matsunaga2014-ar,Matsunaga2013-ck,Matsunaga2012-jb,Cea2016-gv,Xu2019-tt,Watanabe2022-xi,Watanabe2022-yi,Zhang2022-wa,Udina2022-jo,Vaswani2020-pt,Yang2019-ml,Nakamura2020-vj,Lim2020-tn,Zhao2016-sp,Tanaka2022-gx},
may open up a new perspective of materials science which is hardly accessible via conventional experimental techniques.

Among various nonreciprocal phenomena, recent discovery of the superconducting diode effect (SDE)~\cite{Ando2020-om} has brought about an explosion of research works
~\cite{Ando2020-om,Miyasaka2021-ly,Shin2021-je,Lin2021-xr,Bauriedl2022-nq,Narita2022-od,Yuan2022-pz,Daido2022-ox,He2022-px,Ilic2022-kh,Scammell2022-pv,Zhai2022-we,Karabassov2022-xu,Legg2022-ol,Lyu2021-sm,Dobrovolskiy2022-gk,Hou2022-hu,Sundaresh2022-aa,Hope2021-st,Zinkl2021-jl,Jiang2022-lr,Baumgartner2021-lg,Baumgartner2022-gi,Wu2022-ey,Diez-Merida2021-ih,Pal2021-df,Gupta2022-wn,Turini2022-hz,Hu2007-cu,Kopasov2021-xz,Misaki2021-wt,Halterman2022-lc,Zhang2021-ju,Davydova2022-yu,Souto2022-bb,Tanaka2022-qw,Zhang2022-lo}
SDE is the nonreciprocity of the critical current for the phase transition between normal and superconducting states.
When the strength of the applied current is between the leftward and rightward critical currents, electrons flow with forming Cooper pairs in one direction while flow individually in the other, accompanying dissipation.
Such a directionality with zero and finite resistance provides a promising avenue for the future application to diode devices with ultra-low energy consumption.
It has also been pointed out that the SDE offers a promising probe of 
exotic superconducting states with finite center-of-mass momentum of Cooper pairs~\cite{Daido2022-ox,Ilic2022-kh,Lin2021-xr,Scammell2022-pv}.
Thus, further study of the SDE is an urgent issue both from the engineering and fundamental-physics viewpoints.

The SDE of bulk materials was first reported in a superlattice of Nb/V/Ta with Rashba spin-orbit coupling~\cite{Ando2020-om,Miyasaka2021-ly}, where the directionality is controlled by the applied magnetic field. 
Subsequent experiments have revealed the SDE in a NbSe${}_2$ nanowire~\cite{Bauriedl2022-nq} and heterostructure~\cite{Shin2021-je}, and in twisted-trilayer graphene/WSe${}_2$~\cite{Lin2021-xr}.
Therein, it has been demonstrated that the SDE occurs in systems with non-Rashba (Ising- or Zeeman-type) spin-orbit coupling~\cite{Shin2021-je,Bauriedl2022-nq,Lin2021-xr}, and can be triggered by a symmetry-breaking order intertwined with superconductivity~\cite{Lin2021-xr}, instead of the external magnetic field.
The Josephson diode effect, which refers to the SDE realized in Josephson junctions, is also a hot topic~\cite{Baumgartner2022-gi,Turini2022-hz,Gupta2022-wn,Wu2022-ey,Diez-Merida2021-ih,Pal2021-df,Baumgartner2021-lg,Halterman2022-lc,Tanaka2022-qw,Souto2022-bb,Davydova2022-yu,Hu2007-cu,Zhang2021-ju,Misaki2021-wt,Kopasov2021-xz}.
While the diode effect in junction systems has been recognized at least from 2000's~\cite{Reynoso2008-or,Zazunov2009-we,Margaris2010-ax,Yokoyama2014-fw,Silaev2014-bp,Campagnano2015-jl,Dolcini2015-cu,Chen2018-gz,Minutillo2018-dz,Pal2019-ed},
huge nonreciprocity is achieved in recent experiments
~\cite{Baumgartner2022-gi,Turini2022-hz,Gupta2022-wn,Wu2022-ey,Diez-Merida2021-ih,Pal2021-df,Baumgartner2021-lg}.

Theoretical understanding of SDE is still an ongoing issue.
Generally speaking, the critical current of superconductors depends on the sample quality as well as the experimental geometry, in particular when it is related to the vortex motion.
This means that the critical current is extrinsic, and in turn, has a high tunability.
An experiment in a conformal-mapped nanohole system~\cite{Lyu2021-sm} has demonstrated that the nonreciprocity in the flux-flow Joule-heating effect can give rise to a nonreciprocal critical current.
It has also been argued that the different circumstances on sample edges and the associated surface barriers for the vortex entrance lead to the SDE in combination with the Meissner screening current~\cite{Hou2022-hu,Hope2021-st}.
A similar situation occurs in Josephson-coupled two superconducting layers with different superfluid weight~\cite{Sundaresh2022-aa}.
In contrast to the sample-and/or-geometry-sensitive critical currents,
theorists have pointed out that there is an intrinsic nonreciprocity in the critical current of superconducting materials associated with the depairing of Cooper pairs~\cite{Daido2022-ox,Yuan2022-pz,He2022-px,Ilic2022-kh}.
The SDE caused by the depairing mechanism, which is intrinsic to each superconducting material, is called the intrinsic SDE~\cite{Daido2022-ox}, and is expected to be quantitatively feasible especially in small-width samples where the supercurrent flows with a nearly spatially-uniform profile.
It seems unlikely that all the SDE experiments are understood only by the vortex mechanisms, although unambiguous evidence of the SDE by the depairing mechanism has not yet been obtained.
In addition to small-bridge devices~\cite{Bauriedl2022-nq}, field-free setup such as magnetic heterostructures~\cite{Shin2021-je,Narita2022-od} and systems with spontaneous time-reversal-symmetry breaking~\cite{Lin2021-xr} would offer a suitable testground to distinguish the intrinsic SDE from others.
Further study is necessary to uncover the whole picture of the SDE.

An important aspect of the intrinsic SDE is the ability to capture the microscopic properties of the superconducting states.
In particular, those related to the finite center-of-mass momentum of Cooper pairs would directly be probed by the intrinsic SDE, considering the coupling with the supercurrent and the Cooper-pair momentum in the free energy.
Actually, a close relationship with the SDE and the so-called helical superconductivity~\cite{Bauer2012-xi,Smidman2017-hb,Agterberg2003-jn,Barzykin2002-eh,Dimitrova2003-mo,Kaur2005-jf,Agterberg2007-vl,Dimitrova2007-hp,Samokhin2008-nv,Yanase2008-yb,Bauer2012-xi,Michaeli2012-gl,Sekihara2013-dm,Houzet2015-iy}
has been theoretically pointed out~\cite{Daido2022-ox,Ilic2022-kh}.
The SDE has also been used to explore the symmetry breaking of the twisted trilayer graphene/WSe${}_2$, where the finite-momentum pairing induced by the valley polarization has been proposed~\cite{Lin2021-xr,Scammell2022-pv}.

Helical superconductivity
is known as the superconducting state with a spontaneous Cooper-pair momentum, which ubiquitously appears in noncentrosymmetric superconductors under magnetic fields~\cite{Bauer2012-xi,Smidman2017-hb}.
Its appearance is phenomenologically understood by the magnetoelectric coupling of the system.
In analogy with 
magnetoelectric phenomena in normal states,
one might expect that a finite supercurrent flows in the Bardeen-Cooper-Schriefer (BCS) state under a magnetic field.
This is indeed the case, as long as the zero-momentum pairing is assumed to be stabilized even in the presence of the magnetic field~\cite{Yip2002-ao}.
However, we know from a general principle of condensed matter~\cite{Watanabe2019-qh,Ohashi1996-vu,Bohm1949-gk} that a finite-current state does not realize a minimum of the free energy, and a more stable solution can be found by adding small momentum to the system.
Helical superconductivity is thus stabilized under magnetic fields, where the magnetoelectric supercurrent is compensated by the counterflow of finite-momentum Cooper pairs.

Helical superconductivity in Rashba systems shows a characteristic evolution under the magnetic field, with rather different low- and high-field behaviors. 
While the former might be understood as something close to the BCS state as discussed above, the latter is essentially different from the low-field ones and accompanies larger Cooper-pair momenta~\cite{Smidman2017-hb,Bauer2012-xi}.
As increasing the magnetic field, there occurs in the single-$q$ solution either a rapid crossover or a first-order transition between these states depending on the model and the temperature.
It has been shown with the mean-field calculations~\cite{Daido2022-ox} and quasiclassical theory~\cite{Ilic2022-kh} that such a change in the nature of the helical superconductivity accompanies a large diode effect as well as its sign reversal.
Further study of the intrinsic SDE as a probe of helical superconductivity is an important issue, because the experimental implications have been limited to only indirect ones such as the large upper critical field~\cite{Sekihara2013-dm} and the upturn in the temperature-magnetic-field phase diagram~\cite{Schumann2020-mw,Naritsuka2017-or,Naritsuka2021-ym}.

In this paper, we study the supercurrent-related nonreciprocity in superconducting phases, on the basis of both the phenomenological and microscopic arguments.
We discuss in Sec.~\ref{sec:GL} the phenomenological Ginzburg-Landau (GL) theory to derive a general formula for the intrinsic SDE under low magnetic fields.
It is clarified that the effective spin-orbit coupling obtained from the GL coefficients determine the SDE.
The obtained formula offers a convenient criterion to judge whether a finite SDE is obtained for the given anti-symmetric spin-orbit coupling of the system as well as for the given current- and magnetic-field directions.
After showing the formulation of the microscopic study in Sec.~\ref{sec:model_and_setup}, we discuss the SDE of the Rashba-Zeeman model for the $s$-wave and $d$-wave superconductors in Sec.~\ref{sec:SDE_micro}.
It is found that the $d$-wave superconductors show qualitatively similar behaviors, while a larger nonreciprocity tends to be obtained.
In particular, the sign-reversal of the SDE is established as the probe of the crossover of the helical superconductivity regardless of the pairing symmetry.

To further understand the origin of nonreciprocity, in Secs.~\ref{subsec:re-entrantSC} and \ref{subsec:re-entrantSC2}, we study the re-entrant behavior of the critical current.
It is shown that the crossover or the first-order transition occurs in the superconducting solution supporting the critical current.
In Sec.~\ref{sec:critical_field}, we 
discuss the temperature-magnetic-field phase diagram of noncentrosymmetric superconductors under the supercurrent.
We first show in Sec.~\ref{subsec:phenomenology_critical_field} that there is a one-to-one correspondence between the SDE and the nonreciprocity of the transition lines when the transition lines are located where the SDE is small.
The skewness and crossings of the transition lines are explained as the consequence of the SDE.
Such phenomenological results are illustrated with the transition lines of the Rashba-Zeeman superconductors under small or large supercurrent.
On the other hand, such a simple correspondence between the SDE and nonreciprocal transition lines might break down when the SDE is large.
In Sec.~\ref{subsec:critical_field_moderate}, we show that there appears a kink in the transition lines under moderate electric current in Rashba-Zeeman superconductors.
The transition lines can be even re-entrant when they deeply cross the crossover line of the helical superconductivity.
Finally, we make some remarks in Sec.~\ref{sec:discussion} and summarize the paper in Sec.~\ref{sec:summary}.

\section{Phenomenological GL theory}
\label{sec:GL}
In this section, we discuss the intrinsic SDE near the transition temperature.
We extend the results of Refs.~\onlinecite{Daido2022-ox,He2022-px,Yuan2022-pz,Ilic2022-kh} to arbitrary current and magnetic field directions and arbitrary system dimensions $d$.

\subsection{GL free energy for the SDE}
We consider the GL free energy of a noncentrosymmetric superconductor under the magnetic field $\bm{h}$,
\begin{align}
    f(\bm{q},\psi)=\alpha(\bm{q})\psi^2+\frac{\beta(\bm{q})}{2}\psi^4.\label{eq:GL_trivial}
\end{align}
The order parameter has the spatial dependence $\psi e^{i\bm{q}\cdot\bm{x}}$ with the center-of-mass momentum $\bm{q}$.
Assumption of such a single-$q$ order parameter seems to be natural, since the single-$q$ helical state is known to be stabilized near the transition temperature~\cite{Agterberg2007-vl}.
The pairing symmetry is arbitrary as long as it belongs to a one-dimensional representation of the point group.
In the following, we discuss the GL coefficients which should be taken into account to correctly describe the SDE up to $O(h)$, since there seems to be a confusion in the literature~\cite{Daido2022-ox,Yuan2022-pz,He2022-px}.

The GL coefficients are generally written as
\begin{align}
    \alpha(\bm{q})&=\alpha^\zero+\alpha^\one_{i}q_i+\alpha^\two_{ij}q_iq_j\notag\\
    &\quad+\alpha^\three_{ijk}q_iq_jq_k+\alpha^\four_{ijkl}q_iq_jq_kq_l+O(q^5)\label{eq:GLa_original},\\
    \beta(\bm{q})&=\beta^\zero(1+\beta^\one_iq_i+\beta^\two_{ij}q_iq_j)+O(q^3).\label{eq:GLb_original}
\end{align}
The repeated indices $i,j,k,l$ are summed over $i=1,2,\cdots d$.
The GL coefficients are symmetric tensors, and in particular the Lifshitz invariants $\alpha^\one, \alpha^\three,$ and $\beta^\one$ are allowed only in the absence of both the inversion and time-reversal symmetries.
Thus, they are the $O(h)$ quantities.

The free energy is optimized by the order parameter
\begin{align}
    \psi^2=-\frac{\alpha(\bm{q})}{\beta(\bm{q})},
\end{align}
and thus the center-of-mass-momentum dependence of the GL free energy is given by
\begin{align}
    f(\bm{q})=f(\bm{q},\sqrt{-\alpha(\bm{q})/\beta(\bm{q})})=-\frac{\alpha(\bm{q})^2}{2\beta(\bm{q})}.
\end{align}
This is minimized at $\bm{q}=0$ in the absence of the Lifshitz invariants $\alpha^\one,$ $\alpha^\three$, and $\beta^\one$, and thus the BCS state is realized.
On the other hand, their presence leads to the stabilization of the finite-momentum state with $\bm{q}=\bm{q}_0$ given by
\begin{align}
    \partial_{\bm{q}_0}f(\bm{q}_0)=0.
\end{align}
Such a state is called the helical superconductivity, which ubiquitously appears in noncentrosymmetric superconductors under magnetic fields.

To discuss the SDE up to $O(h)$, it is convenient to trace out the $q$-linear term in $\alpha(\bm{q})$.
This is achieved by shifting $\bm{q}$ by the solution of $\partial_{q_i}\alpha(\bm{q})=0$, which is written as $\tilde{\bm{q}}_0$.
We obtain
\begin{align}
    \tilde{q}_{0i}=-\frac{1}{2}[\alpha^\two]_{ij}^{-1}\alpha_j^\one+O(h^2),
\end{align}
which coincides with $\bm{q}_0$ within the standard GL theory where the higher-order GL cofficients are neglected.
We obtain for $\delta\bm{q}=\bm{q}-\tilde{\bm{q}}_0$,
\begin{align}
    \alpha(\bm{q}_0+\delta\bm{q})&=\alpha_0+[\alpha_2]_{ij}\delta q_i\delta q_j\notag\\
    &\qquad+[\alpha_3]_{ijk}\delta q_i\delta q_j\delta q_k+O(\delta q^4,h^2),\label{eq:GLa}\\
    \beta(\bm{q}_0+\delta\bm{q})&=\beta_0(1+[\beta_1]_i\delta q_i)+O(\delta q^2,h^2),\label{eq:GLb}
\end{align}
with coefficients
\begin{align}
\alpha_0&=\alpha^\zero,\quad
[\alpha_2]_{ij}=\alpha^\two_{ij},\quad
\beta_0=\beta^\zero,
\end{align}
as well as
\begin{align}
[\alpha_3]_{ijk}
&=\alpha^\three_{ijk}-2\alpha^\four_{ijkl}[\alpha^\two]^{-1}_{lm}\alpha_m^\one,\label{eq:alpha_3_GL}\\
[\beta_1]_i
&=\beta^\one_i-\beta^{\two}_{ij}[\alpha^\two]^{-1}_{jk}\alpha_l^\one,\label{eq:beta_1_GL}
\end{align}
neglecting all the $O(h^2)$ contributions such as $\tilde{q}_0^2$ and $\tilde{q}_0\alpha^\three$.
Equations~\eqref{eq:GLa} and \eqref{eq:GLb} are exact for the description up to $O(h)$, and are the natural generalization of the GL free energy studied in Ref.~\onlinecite{Daido2022-ox}.

Let us see that all the relevant terms are included to the GL free energy from the perspective of the temperature scaling~\cite{Daido2022-ox,He2022-px}. 
Note that the free energy $f(\bm{q},\psi)$ with GL coefficients in Eqs.~\eqref{eq:GLa} and \eqref{eq:GLb} includes terms of $O(\delta q^n\psi^m)$ up to $n+m\le 5$.
Since the normal-state transition occurs for $\delta q\sim \sqrt{T_{\rm c}-T}$
\footnote{This follows from the rough estimate of the region where $f(\bm{q})\sim -(T_{\rm c}-T)+\delta q^2<0$, neglecting the higher-order corrections.
To be precise, it is more appropriate to understand $\delta q$ here as $\bm{q}-\bm{q}_0$ rather than $\bm{q}-\tilde{\bm{q}}_0$.
However, $\tilde{\bm{q}}_0-\bm{q}_0=O(h(T_{\rm c}-T))$, and their difference does not affect the discussion.
See also Appendix~\ref{app:GL_general_h} for this point.
},
which scales with the inverse of the correlation length,
we are interested in the $q$ range $|\delta q|\lesssim \sqrt{T_{\rm c}-T}$.
Considering that $\psi\sim\sqrt{T_{\rm c}-T}$, the free energy takes into account all the terms up to $O(T_{\rm c}-T)^{5/2}$.
This allows us to correctly describe the electric current $\bm{j}(\bm{q})=2\partial_{\bm{q}}f(\bm{q})\sim O(f(\bm{q})/\sqrt{T_{\rm c}-T})$ up to $O(T_{\rm c}-T)^2$, which is sufficient to consider the SDE of $O(h)$ as clarified in the following.
It should also be noted that $q\sim q_0+O(\sqrt{T_{\rm c}-T})$, and therefore, $\alpha^\four q^4\psi^2$ and $\beta^\two q^2\psi^4$ includes the contribution of the order $h(T_{\rm c}-T)^{5/2}$.
This is the reason why $\alpha^\four$ and $\beta^\two$ terms should be kept in Eqs.~\eqref{eq:GLa_original} and \eqref{eq:GLb_original} while $\delta q^4\psi^2$ and $\delta q^2\psi^4$ terms can be neglected in Eqs.~\eqref{eq:GLa} and \eqref{eq:GLb}.
Note that $\psi^6\sim (T_{\rm c}-T)^3$ can also be neglected.

Before proceeding, we simplify the notations of the GL coefficients.
We can always choose the coordinate axes to diagonalize the real symmetric matrix $\alpha_2=\alpha^\two$.
We choose such a coordinate system in the following and write
\begin{align}
    [\alpha_2]_{ij}=\alpha^\two_{ij}=\frac{1}{2m_i}\delta_{ij}.
\end{align}
Note that $m_i>0$ by naturally assuming that the BCS state is the most stable for $\bm{h}=0$.
Since $\alpha_3$ and $\beta_1$ are proportional to the magnetic field $\bm{h}$, we can write
\begin{align}
    [\alpha_3]_{ijk}\delta q_i\delta q_j\delta q_k&\equiv\bm{h}\cdot\bm{g}_3(\delta\bm{q}),\\
    [\beta_1]_{ij}\delta q_i&\equiv \bm{h}\cdot\bm{g}_1(\delta\bm{q}),
\end{align}
where the functions $\bm{g}_3(\delta\bm{q})$ and $\bm{g}_1(\delta\bm{q})$ are homogeneous polynomials of degree three and one, obtained from Eqs.~\eqref{eq:alpha_3_GL} and \eqref{eq:beta_1_GL}, respectively.
Note also that the coefficient $\alpha_0$ is proportional to $T-T_{\rm c}$ up to $O(h^2)$, and can be written as $\alpha_0=-a_0(T_{\rm c}-T)$ with $a_0>0$.
Finally, we arrive at the GL free energy for the SDE up to $O(h)$, that is, Eq.~\eqref{eq:GL_trivial} with
\begin{align}
    \alpha(\bm{q})&=-a_0(T_{\rm c}-T)+\sum_i\frac{\delta q_i^2}{2m_i}+\bm{g}_3(\delta\bm{q})\cdot\bm{h},\label{eq:GLc}\\
    \beta(\bm{q})&=\beta_0(1+\bm{g}_1(\delta\bm{q})\cdot\bm{h}).
\end{align}

\subsection{Symmetry of the GL coefficients}
Note that the diode effect vanishes in the absence of the coefficients $\bm{g}_3$ and $\bm{g}_1$~\cite{Smidman2017-hb}.
\footnote{
In contrast to Refs.~\onlinecite{Bauer2012-xi,Smidman2017-hb}, Edelstein concluded the finite SDE by the depairing mechanism, within the GL theory taking into account only the first-order Lifshitz invariant.~\cite{Edelstein1996-ro}
It seems that the magnetization current unphysically contributes to the net current in Ref.~\onlinecite{Edelstein1996-ro}.
Our treatment agrees with Refs.~\onlinecite{Bauer2012-xi,Smidman2017-hb}.}
Actually, {for $\bm{g}_1 = \bm{g}_3 =0$} the GL free energy of the system is equivalent with that of a BCS superconductor except for the origin of the momentum, whose shift does not affect the depairing critical current.
Thus, these coefficients are essential for the diode effect,
and their symmetry properties are commented in the following.

It should be noted that 
$\bm{g}_3(\bm{q})\cdot\bm{h}$ and $\bm{g}_1(\bm{q})\cdot\bm{h}$ remain invariant against the point-group operations simultaneously on $\bm{q}$ and $\bm{h}$, as they are included in the free energy
(Note that $\bm{q}$ and $\delta\bm{q}$ behave in the same way for the point-group operations of the system).
This means that $\bm{g}_3(\bm{k})\cdot\bm{\sigma}$ and $\bm{g}_1(\bm{k})\cdot\bm{\sigma}$ are allowed to appear in the Bloch Hamiltonian from symmetry points of view, and vice versa, since the magnetic field $\bm{h}$ and the spin $\bm{\sigma}$ behave in the same way. 
Thus, the GL coefficients $\bm{g}_3$ and $\bm{g}_1$ are symmetry-equivalent with the anti-symmetric spin-orbit coupling (ASOC) of $O(k^3)$ and $O(k)$, respectively.

Out of 21 noncentrosymmetric point groups,
$k$-linear ASOC is allowed in 18 ones (the gyrotropic point groups~\cite{He2020-yq}), while it is forbidden in $T_d$, $D_{3h}$, and $C_{3h}$.
For polar point groups, 
for example, 
the Rashba spin-orbit coupling such as $\bm{g}_1(\bm{k})\sim (-k_y,k_x,0)$ exists. 
On the other hand, the third-order ASOC
is allowed in all the noncentrosymmetric point groups, and in particular, are known as the Dresselhaus spin-orbit coupling
\begin{align}
\bm{g}_3(\bm{k})\sim(k_xk_y^2,-k_yk_x^2,0),    
\end{align}
for $T_d$ and as Ising or Zeeman spin-orbit coupling such as 
\begin{align}
\bm{g}_3(\bm{k})\sim(0,0,k_y^3-3k_yk_x^2),
\end{align}
for $D_{3h}$ and $C_{3h}$. Here, $\bm{g}_3(k_z=0)$ is shown for simplicity. 
For the gyrotropic point groups, $\bm{g}_3(\bm{k})$ can include $\bm{k}^2\bm{g}_1(\bm{k})$, for example, where $\bm{k}^2$ can be replaced with an arbitrary $O(k^2)$ term belonging to the identity representation.

In summary, the Lifshitz invariants $\bm{g}_1(\bm{q})$ and $\bm{g}_3(\bm{q})$ have the same symmetry property as the ASOC characteristic of each noncentrosymmetric point group; therefore, their wave-number dependence is similar to that of the ASOC near the Gamma point {in the Brillouin zone}.
For explicit functional forms of $\bm{g}_1(\bm{q})$ and $\bm{g}_3(\bm{q})$, see the appendix of Ref.~\onlinecite{Frigeri2005-ls} showing the classification of the ASOC.

\subsection{GL formula for the SDE}
Let us derive the SDE up to $O(h)$.
This can be achieved by maximizing and minimizing the current
\begin{align}
    \bm{j}(\bm{q})&=2\partial_{\bm{q}}f(\bm{q}),
\end{align}
to obtain the critical current
\begin{align}
    j_{{\rm c} +}(\hat{n})=\max_{\bm{q}}[\hat{n}\cdot\bm{j}(\bm{q})],\quad j_{{\rm c} -}(\hat{n})=\min_{\bm{q}}[\hat{n}\cdot\bm{j}(\bm{q})].\label{eq:def_jc}
\end{align}
Here, the direction of the electric current is chosen to be parallel or antiparallel to the unit vector $\hat{n}$.
Accordingly, the nonreciprocity in the depairing critical current is obtained as
\begin{align}
\Delta j_{\rm c}(\hat{n})\equiv j_{{\rm c} +}(\hat{n})-|j_{{\rm c} -}(\hat{n})|=j_{{\rm c} +}(\hat{n})+j_{{\rm c} -}(\hat{n}).\label{eq:def_Djc}
\end{align}
The intrinsic SDE means that $\Delta j_{\rm c}$ takes a finite value.
We also define the averaged critical current,
\begin{align}
    \bar{j}_{\rm c}(\hat{n})\equiv\frac{1}{2}\left[j_{{\rm c} +}(\hat{n})+|j_{{\rm c} -}(\hat{n})|\right],\label{eq:def_jcbar}
\end{align}
by which the diode quality factor is defined by
\begin{align}
    r(\hat{n})&=\frac{j_{{\rm c} +}(\hat{n})-|j_{{\rm c} -}(\hat{n})|}{j_{{\rm c} +}(\hat{n})+|j_{{\rm c} -}(\hat{n})|}=\frac{\Delta j_{\rm c}(\hat{n})}{2\bar{j}_{\rm c}(\hat{n})}.\label{eq:def_r}
\end{align}
This quantifies the degree of nonreciprocity.

The SDE is obtained
by studying the first-order perturbation to $j_{{\rm c} \pm}(\hat{n})$ by $\bm{g}_3$ and $\bm{g}_1$.
The calculation is done in a way similar to Ref.~\onlinecite{Daido2022-ox}, and the details are given in Appendix~\ref{app:GL_derivation}.
We obtain
\begin{align}
\Delta j_{\rm c}(\hat{n})&=\frac{8a_0^2}{9\beta_0}(T_{\rm c}-T)^2\,\bm{g}_{\mathrm{eff}}(\hat{n})\cdot\bm{h},  \label{eq:Djc_GL}\\
{r}(\hat{n})
&=\sqrt{\frac{a_0m(\hat{n})(T_{\rm c}-T)}{6}}\ \bm{g}_{\mathrm{eff}}(\hat{n})\cdot\bm{h}, \label{eq:r_GL}
\end{align}
up to $O(h)$.
Here, the effective ASOC $\bm{g}_{\mathrm{eff}}(\bm{q})$ for the diode effect is defined by
\begin{equation}
    \bm{g}_{\mathrm{eff}}(\bm{q})\equiv\frac{2\bm{g}_3(\bm{q})}{\displaystyle\sum_i{q_i^2}/{2m_i}}-\bm{g}_1(\bm{q}),
\label{eq:g_eff}
\end{equation}
while $1/m(\hat{n})\equiv\sum_i\hat{n}_i^2/m_i$.
{Equations~\eqref{eq:Djc_GL}-\eqref{eq:g_eff} are one of the central results of this section.}
The symmetry of $\bm{g}_{\mathrm{eff}}(\bm{q})$ is equivalent to that of the ASOC of the system, since $\sum_iq_i^2/2m_i$ belongs to the identity representation of the point group.
Therefore, we can replace $\bm{g}_{\mathrm{eff}}(\bm{q})$ with the spin-orbit coupling of the system $\bm{g}(\bm{k})$ for the purpose of symmetry considerations, giving a convenient criterion to obtain the SDE.
{For a quantitative estimation, we have to evaluate $\bm{g}_{\mathrm{eff}}(\bm{q})$ with Eq.~\eqref{eq:g_eff}.}

Equation~\eqref{eq:Djc_GL} reduces to the result of Ref.~\onlinecite{Daido2022-ox} for  2D Rashba systems with $\hat{n}=\hat{x}$ and $\bm{h}=h\hat{y}$.
According to the formula~\eqref{eq:Djc_GL}, $\Delta j_{\rm c}$ and $r$ are proportional to $(T_{\rm c}-T)^2$ and $\sqrt{T_{\rm c}-T}$, respectively, which are consistent with the reciprocal component of the critical current $\bar{j}_{\rm c}\sim (T_{\rm c}-T)^{3/2}$~\cite{Tinkham2004-dh}.
While we have focused on the SDE of $O(h)$, the temperature scaling $\Delta j_{\rm c}\propto(T_{\rm c}-T)^2$ and $r\propto\sqrt{T_{\rm c}-T}$ hold even when higher-order effects of the magnetic field are taken into account [See Appendix~\ref{app:GL_general_h} for details], and it is a general feature of the intrinsic SDE.
The temperature scaling has been found in Refs.~\onlinecite{Daido2022-ox,He2022-px,Yuan2022-pz}
\footnote{The temperature scaling of the nonreciprocity has also been commented in the old literature without an explicit calculation~\cite{Levitov1985-pm}.},
and has been confirmed with mean-field calculations~\cite{Daido2022-ox,He2022-px}.
The result $\Delta j_{\rm c}(\hat{n})\propto (T_{\rm c}-T)^2$ is intuitive, since
$\Delta j_{\rm c}$ is caused by $\bm{g}_3$ and $\bm{g}_1$, both of which are $O(T_{\rm c}-T)^{5/2}$ terms in $f(\bm{q})$, and naturally give rise to electric current of the order $(T_{\rm c}-T)^{5/2}/\sqrt{T_{\rm c}-T}=(T_{\rm c}-T)^2$.

\subsection{Discussion}
The intrinsic SDE is contributed not only by the cubic term $\alpha_3$ in $\alpha(\bm{q})$ but also by the linear term $\beta_1$ in $\beta(\bm{q})$.
This was first pointed out in Ref.~\onlinecite{Daido2022-ox}, while the renormalization of the coefficients $\alpha_3$ and $\beta_1$ by $\alpha^\four$ and $\beta^\two$ was overlooked
and has later been pointed out in Refs.~\onlinecite{Ilic2022-kh,He2022-px}\footnote{
$\beta^\one$ and $\beta^\two$ are additionally included in the latest version of the preprint as well as the published paper of Ref.~\onlinecite{He2022-px}, while were not in previous versions. 
In this paper, we refer to Ref.~\onlinecite{He2022-px} as its published version.}.
The importance of $\beta_1$ (or an equivalent quantity) has been emphasized in Ref.~\onlinecite{Ilic2022-kh}, and it has been shown that the $O(h)$ SDE vanishes in the ideally isotropic 2D Rashba $s$-wave superconductor near the transition temperature, due to the cancellation between the contributions from $\alpha_3$ and $\beta_1$.
It has also been pointed out that the forbidden $O(h)$ SDE of the isotropic Rashba model is obtained in Ref.~\onlinecite{Yuan2022-pz} 
because $\beta_1$ is not taken into account~\cite{Ilic2022-kh}.
\footnote{On the other hand, Ref.~\onlinecite{He2022-px} reports $O(h)$ SDE in the ideally isotropic Rashba model with chemical potential near the Dirac point by a mean-field calculation.
The reason for the appearance of $O(h)$ SDE might be the effects beyond the quasiclassical approximation.
In any case, neglecting $\beta_1$ is not quantitatively justified.}
On the other hand, such a cancellation of $\alpha_3$ and $\beta_1$ is due to the simpleness of the isotropic Rashba model, and generally the $O(h)$ SDE exists.
It is expected that the anisotropy of systems, including that of the order parameter, is important to obtain a large $\bm{g}_{\mathrm{eff}}(\bm{q})$ and the large $O(h)$ SDE, while the importance of the anisotropy is also manifested for the low-temperature SDE governed by the nonreciprocity of the Landau critical momentum measured from $q_0$~\cite{Daido2022-ox}.

Let us draw from Eq.~\eqref{eq:Djc_GL} the condition to realize the intrinsic SDE.
First, the direction $\hat{n}$ of the electric current must be chosen so that the ASOC becomes finite in that direction, i.e. $\bm{g}_{\mathrm{eff}}(\hat{n})\neq0$.
In particular, the intrinsic SDE is not obtained when the electric current is applied along the high-symmetry lines of $T_d$ systems ($D_{3h}$ and $C_{3h}$ systems) where
the Dresselhaus (Ising or Zeeman) ASOC identically vanishes~\cite{Yuan2022-pz,He2022-px}.
In this way, $\hat{n}$ dependence, namely the current-direction dependence with respect to the crystal axes, generally follows from that of the effective spin-orbit coupling $\bm{g}_{\mathrm{eff}}(\hat{n})$ as well as $m(\hat{n})$ for the quality factor $r(\hat{n})$.
In addition, the magnetic field must have a component parallel to the ASOC.
When the direction of $\bm{h}$ is rotated with fixing the current direction $\hat{n}$, the angle dependence is given by $\cos\theta$, with $\theta$ the relative angle between the vectors $\bm{h}$ and $\bm{g}_{\mathrm{eff}}(\hat{n})$.
Such a one-fold angle dependence has also been observed for the magnetochiral  anisotropy~\cite{Ideue2017-vp,Ideue2021-es}.
Higher harmonics may also appear when higher-order corrections of $\bm{h}$ are taken into account.

In closing this section, we make a comment on the relation of the SDE with the helical superconductivity.
Helical superconductivity is realized by the $q$-linear term in $f(\bm{q})$.
Thus, helical superconductivity, meaning finite-momentum superconductivity in equilibrium, is realized only in gyrotropic point groups at least within the GL theory under low magnetic fields.
{On the other hand,} it should be noted that $\bm{g}_3(\bm{q})$ is generally finite even in non-gyrotropic point groups $T_d$, $D_{3h}$ and $C_{3h}$.
Therefore, the SDE is allowed in all the 21 noncentrosymmetric point groups and not restricted to the gyrotropic ones,
in agreement with the 
observation of SDE in trigonal crystal structure~\cite{Shin2021-je,Bauriedl2022-nq}.
As we have seen, the SDE occurs by the asymmetry of $f(\bm{q})$ around $\bm{q}_0$, rather than the finite equilibrium momentum $\bm{q}_0\neq0$ itself.
In this sense, the SDE is not directly related to the helical superconductivity.
Nevertheless, the SDE captures the non-perturbative information of the momentum dependence of the condensation energy, and thereby detects the characteristic crossover of the helical superconductivity in gyrotropic systems.
Thus, the relationship with the helical superconductivity is clarified only by correctly introducing the nonlinear effects of the magnetic fields, and is beyond the phenomenological GL theory.
Microscopic studies such as Bogoliubov-de Gennes~\cite{Daido2022-ox} and quasiclassical~\cite{Ilic2022-kh} mean-field theories, as well as the GL theory with coefficients determined by them,
are suitable to describe these nonlinear effects.
In the following sections, we discuss the mean-field theory for the intrinsic SDE.

\section{Microscopic study of the depairing critical current in Rashba-Zeeman model}
In this section, we microscopically study the nonreciprocity triggered by the supercurrent.
We focus on the $s$-wave and $d$-wave superconductivity in the Rashba-Zeeman model.
After showing the model and setup in Sec.~\ref{sec:model_and_setup}, we reproduce in Sec.~\ref{sec:SDE_micro} the SDE for $s$-wave superconductors, and compare them with the results for the $d$-wave superconductors.
Furthermore, we discuss the re-entrant behavior in the critical current from the microscopic viewpoint in Sec.~\ref{subsec:re-entrantSC}, and also discuss the characteristic first-order transition and crossover of superconducting states in Sec.~\ref{subsec:re-entrantSC2}.

\subsection{Model and setup}
\label{sec:model_and_setup}
We show the model to discuss the nonreciprocity in the $s$-wave and $d$-wave superconductors. 
Following the setup of Ref.~\onlinecite{Daido2022-ox}, we consider the Rashba-Zeeman model with an attractive interaction,
\begin{align}
\hat{H}&=\sum_{\bm k \sigma\sigma'}\bigl[\xi (\bm k)\delta_{\sigma\sigma'}
+\{\bm{g}(\bm k)-\bm{h}\} \cdot \bm{\sigma}_{\sigma \sigma'}]
c_{\bm k \sigma}^\dagger c_{\bm k \sigma'} 
\label{eq:model}
\\
&\quad-\sum_{\bm{k},\bm{k}',\bm{q}'}c^\dagger_{\bm{k}+\bm{q}',a}c^\dagger_{-\bm{k}+\bm{q}',b}U_{abcd}(\bm{k},\bm{k}',\bm{q}')c_{-\bm{k}'+\bm{q}',d}c_{\bm{k}'+\bm{q}',c}.\notag \label{eq:interaction}
\end{align}
Note that the definition of $\bm{q}$ is different from Ref.~\onlinecite{Daido2022-ox} and the previous section by a factor of 2, for convenience.
We define the normal-state Bloch Hamiltonian $H_N(\bm{k})_{\sigma\sigma'}$ by the square bracket in the first line of Eq.~\eqref{eq:model}, which contains the hopping term
\begin{align}
\xi(\bm{k})=-2t_1(\cos k_x+\cos k_y)
-\mu,
\end{align}
the Rashba spin-orbit coupling
\begin{align}
    \bm{g}(\bm{k})=\alpha_{\rm g}(-\sin k_y,\sin k_x,0),
\end{align}
and the inplane magnetic field $\bm{h}$.
The attractive interaction 
\begin{align}
    U_{abcd}(\bm{k},\bm{k}',\bm{q}')=\frac{U}{2V}\varphi_{ab}(\bm{k})\varphi_{dc}^\dagger(\bm{k}')\delta_{\bm{q},\bm{q}'},
\end{align}
describes the pairing channel $\varphi_{ab}(\bm{k})$ with a finite center-of-mass momentum $\bm{q}$.
We focus on the $s$-wave and $d$-wave symmetries,
\begin{align}
    \varphi(\bm{k})&=\begin{cases}i\sigma_y&(s\text{-wave})\\
(\cos k_x-\cos k_y)i\sigma_y&(d\text{-wave})
\end{cases}.
\end{align}

\begin{figure}
    \centering
    \includegraphics[width=0.49\textwidth]{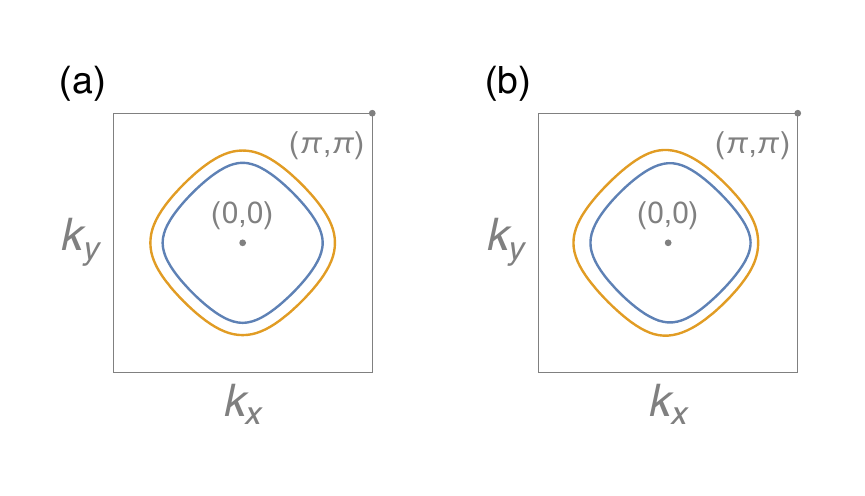}
    \caption{Fermi surfaces of the model for (a) $\bm{h}=0$, and (b) $\bm{h}=0.1\,\hat{y}$.
    The inner (outer) Fermi surface is shifted in the $+\hat{x}$ ($-\hat{x}$) direction by the application of the magnetic field.
    }
    \label{fig:FS}
\end{figure}

The model parameters used in numerical calculations are given as follows:
\begin{align}
    (t_1,\mu,\alpha_{\rm g})=(1,-1,0.3),
\end{align}
which are the same as those in Ref.~\onlinecite{Daido2022-ox}.
The strength of the attractive interaction $U$ is given by
\begin{align}
    U^s=0.75,\quad \text{and}\quad U^{d}=0.6,
\end{align}
for $s$-wave and $d$-wave symmetries, respectively.
They are chosen so as to give similar transition temperatures {$T_{\rm c}\sim0.04$ in units of $t=1$. The qualitative behaviors such as the sign reversals of the SDE are almost independent of the transition temperature, and similar phase diagrams are obtained when $h$ and $T$ are scaled by $T_{\rm c}$.}
Note that the value of $U^s$ is equivalent to $U=1.5$ of Ref.~\onlinecite{Daido2022-ox}, since the notation is changed by a factor of two.
Thus, the $s$-wave state studied in this paper is equivalent to that of Ref.~\onlinecite{Daido2022-ox} without the next-nearest-neighbour hopping.

The Fermi surfaces of the model is shown in Fig.~\ref{fig:FS}(a).
By applying the inplane magnetic field, the inner and outer Fermi surfaces are shifted in the right and left directions, respectively~\cite{Smidman2017-hb}, as show in Fig.~\ref{fig:FS}(b) for $h=0.1$.
Actually, the energy dispersion under $h$ is given by, for the band with helicity $\chi=\pm1$,
\begin{align}
    \epsilon_\chi(\bm{k},\bm{h})&=\xi(\bm{k})+\chi|\bm{g}(\bm{k})-\bm{h}|\notag\\
    &\simeq\xi(\bm{k})+\chi g(\bm{k})-\chi\hat{g}(\bm{k})\cdot\bm{h}\notag\\
    &=\epsilon_\chi(\bm{k},0)-\frac{[\chi\hat{g}(\bm{k})\cdot\bm{h}]\nabla_{\bm{k}}\epsilon_\chi(\bm{k},0)}{|\nabla_{\bm{k}}\epsilon_\chi(\bm{k},0)|^2}\cdot\nabla_{\bm{k}}\epsilon_\chi(\bm{k},0)\notag\\
    &\simeq\epsilon_\chi(\bm{k}-\chi\delta\bm{q}(\bm{k},\bm{h}),0).
\end{align}
Here, we defined $g(\bm{k})\equiv|\bm{g}(\bm{k})|$, $\hat{g}(\bm{k})=\bm{g}(\bm{k})/g(\bm{k})$, and 
\begin{align}
    \delta\bm{q}(\bm{k},\bm{h})\equiv\frac{\hat{g}(\bm{k})\cdot\bm{h}}{|\nabla_{\bm{k}}\epsilon_\chi(\bm{k},0)|^2}\nabla_{\bm{k}}\epsilon_\chi(\bm{k},0).
\end{align}
According to the odd $\bm{k}$-parity of both $\hat{g}(\bm{k})$ and $\nabla_{\bm{k}}\epsilon_\chi(\bm{k},0)$, as well as from the gyrotorpic point-group symmetry,
Fermi-surface average of $\delta\bm{q}(\bm{k},\bm{h})$ takes a finite value $\delta\bm{q}$, which is of the order $h/v_F$ with $v_F$ the Fermi velocity.
This leads to the helicity-dependent shift of the Fermi surfaces $0=\epsilon_\chi(\bm{k},\bm{h})\sim\epsilon_\chi(\bm{k}-\chi\delta\bm{q},0)$.

We evaluate the depairing critical current with the mean-field approximation.
The pair potential $\Delta(\bm{q})$ corresponding to the pairing symmetry $\varphi(\bm{k})$ is introduced by
\begin{equation}
    \sum_{\bm{k}}\Delta(\bm{q})\varphi_{ab}(\bm{k})c^\dagger_{\bm{k}+\bm{q},a}c^\dagger_{-\bm{k}+\bm{q},b}+\text{H.c.}+\Delta(\bm{q})^2/2U,
\end{equation}
approximating the interaction term of the Hamiltonian, Eq.~\eqref{eq:interaction}.
Here, we choose the phase of the order parameter as $\Delta(\bm{q})\ge0$.
Thus, the mean-field Hamiltonian reads
\begin{align}
    \hat{H}_{\mathrm{MF}}(\bm{q})&=\frac{1}{2}\sum_{\bm{k}}\Psi_{\bm{q}}^\dagger(\bm{k})H_{\bm{q}}(\bm{k})\Psi_{\bm{q}}(\bm{k})\notag\\
    &\quad+\frac{1}{2}\sum_{\bm{k}\sigma}\left[[H_N(\bm{k})]_{\sigma\sigma}+\Delta(\bm{q})^2/2U\right],
\end{align}
with the Bogoliubov-de Gennes (BdG) Hamiltonian
\begin{align}
        H_{\bm{q}}(\bm{k})=\begin{pmatrix}
H_N(\bm{k}+\bm{q})&\Delta({\bm{q}})\varphi(\bm{k})\\\Delta({\bm{q}})\varphi(\bm{k})^\dagger&-H_N^T(-\bm{k}+\bm{q})
\end{pmatrix},
\end{align}
and the Nambu spinor 
\begin{align}
\Psi^\dagger_{{\bm{q}}}(\bm{k})\equiv(c_{\bm{k}+\bm{q},\uparrow}^\dagger,c_{\bm{k}+\bm{q},\downarrow}^\dagger,c_{-\bm{k}+\bm{q},\uparrow},c_{-\bm{k}+\bm{q},\downarrow}).
\end{align}
The order parameter $\Delta(\bm{q})$ is determined self-consistently by the gap equation,
\begin{align}
    \Delta({\bm{q}})
    &=-\frac{U}{V}\sum_{\bm{k},n}\braket{u_{n,{\bm{q}}}(\bm{k})|\tau_-\varphi(\bm{k})^\dagger|u_{n,{\bm{q}}}(\bm{k})}f(E_{n,{\bm{q}}}(\bm{k})).
\end{align}
Here, $f(E)\equiv (e^{E/T}+1)^{-1}$ is the Fermi distribution function with the temperature $T$. $\tau_\mu$ represents the Pauli matrices in the Nambu space, and $\tau_\pm\equiv (\tau_x\pm i\tau_y)/2$.
The eigenstates and eigenvalues of the BdG Hamiltonian are defined by
\begin{align}
    H_{{\bm{q}}}(\bm{k})\ket{u_{n,{\bm{q}}}(\bm{k})}=E_{n,{\bm{q}}}(\bm{k})\ket{u_{n,{\bm{q}}}(\bm{k})}.
\end{align}

To obtain the critical current, we need to know the threshold value of $|\bm{j}|$ above and below which no superconducting solutions are obtained.
In doing so, we introduce the function $\bm{j}({\bm{q}})$, which translates the momentum ${\bm{q}}$, a parameter in the Hamiltonian, with the electric current $\bm{j}$.
By using the solution of the gap equation, the electric current for a given $\bm{q}$ can be calculated by
\begin{align}
    \bm{j}(\bm{q})&=\frac{1}{2V}\sum_{n,\bm{k}}\braket{u_{n,\bm{q}}(\bm{k})|\bm{V}_{\bm{q}}(\bm{k})|u_{n,\bm{q}}(\bm{k})}f(E_{n,\bm{q}}(\bm{k})),
\end{align}
with the matrix\footnote{When the separable interaction adopted in this paper is taken seriously, the current operator includes contribution from the interaction term. We neglect such contributions for simplicity.
This corresponds to an implicit assumption that the pairing interaction is obtained from some microscopic Hamiltonian respecting the local $U(1)$ symmetry (and therefore the current operator is determined only by the one-body part), and mixing between pairing channels is negligible for the physical or symmetry reasons.
}
\begin{align}
    \bm{V}_{\bm{q}}(\bm{k})&=\begin{pmatrix}\bm{v}(\bm{k}+\bm{q})&0\\0&-\bm{v}(-\bm{k}+\bm{q})^T\end{pmatrix}_\tau,\\
    \bm{v}(\bm{k})&\equiv\partial_{\bm{k}}H_N(\bm{k}).
\end{align}
When we consider the electric current in the direction parallel and antiparallel to the unit vector $\hat{n}$,
the depairing critical currents and related quantities are obtained by Eqs.~\eqref{eq:def_jc}, \eqref{eq:def_Djc}, \eqref{eq:def_jcbar}, and \eqref{eq:def_r}.
In principle, maximization and minimization of $\hat{n}\cdot\bm{j}(\bm{q})$ to obtain the critical currents should be done for both $q_x$ and $q_y$.
However, the problem might be simplified when $\hat{n}$ and the magnetic field are aligned to high-symmetry axes.
In this paper, we consider the situation
\begin{align}
    \hat{n}=\hat{x},\quad \bm{h}=h\hat{y}.
\end{align}
In this case, it is natural to consider the variation of $\bm{q}$ within the form 
\begin{align}
\bm{q}=q\hat{x}.
\end{align}
This is because $q_y=0$ is a solution of maximization/minimization with respect to $\bm{q}$ owing to the $y$ mirror symmetry, and it is also physically expected to be most favorable in our model.

It is convenient to introduce the condensation energy to discuss the nature of the superconducting state.
The free energy density $\Omega({q},\Delta(q))$ is given by
\begin{align}
    \Omega(q,\Delta(q))&=\frac{1}{2V}\sum_{\bm{k}\sigma}\left[[H_N(\bm{k})]_{\sigma\sigma}+|\Delta(q)|^2/2U\right]\\
    &\quad-\frac{T}{2V}\sum_{\bm{k},n}\ln(1+e^{- E_{n,q}(\bm{k})/T}).
\end{align}
Thereby, the condensation energy is obtained as
\begin{align}
    F(q)\equiv \Omega(q,\Delta(q))-\Omega(q,0),\label{eq:F_of_q}
\end{align}
which is connected to $j(q)$ via $j(q)=\partial_qF(q)$~\cite{Daido2022-ox}.
The condensation energy $F(q)$ can be identified with the Ginzbrug-Landau free energy $f(q)$ near the transition temperature.
The Cooper-pair momentum $q_0$ of the helical superconductivity is obtained by minimizing the condensation energy,
\begin{align}
    F(q_0)\equiv \min_{q}F(q).\label{eq:def_q0}
\end{align}

\subsection{SDE in $s$-wave and $d$-wave superconductors}
\label{sec:SDE_micro}

\begin{figure}[t]
    \centering
    \begin{tabular}{ll}
    (a)  $s$-wave state&(b)\\
        \includegraphics[width=0.25\textwidth]{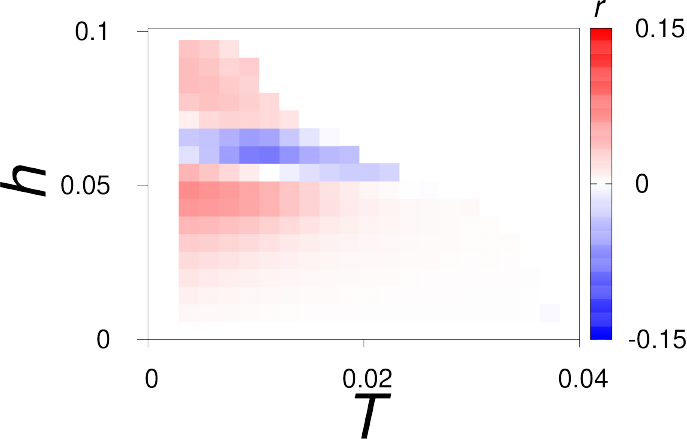}
    &\includegraphics[width=0.25\textwidth]{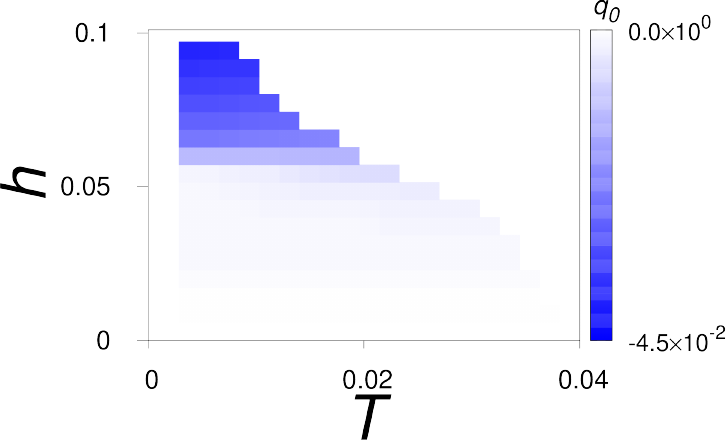}\\
        (c)  $d$-wave state&(d)\\
    \includegraphics[width=0.25\textwidth]{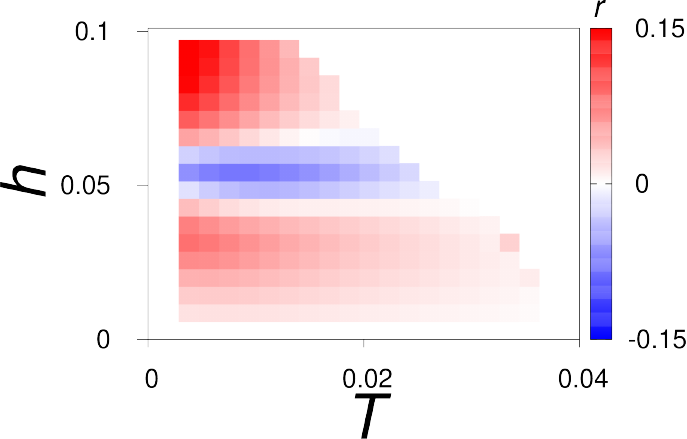}
    &\includegraphics[width=0.25\textwidth]{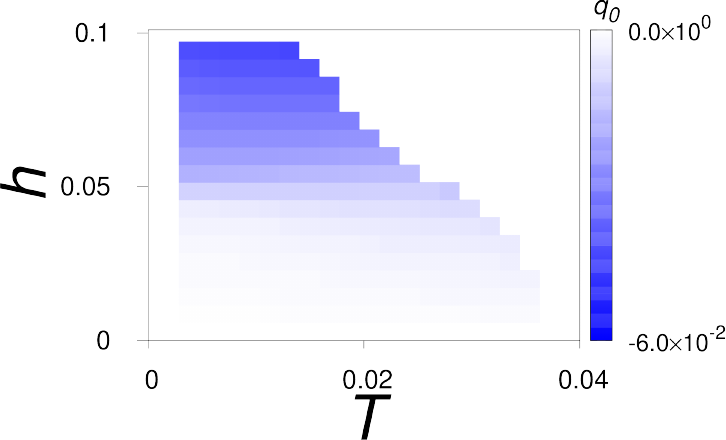}
    \end{tabular}
    \caption{Temperature-magnetic-field phase diagram for the diode quality factor $r(T,h)$ [panels (a) and (c)] and the most stable Cooper-pair momentum $q_0(T,h)$ [panels (b) and (d)] for (a),(b) $s$-wave and (c),(d) $d$-wave states.
    We adopted $L_x=6000$ and $L_y=200$
    for numerical calculations.}
    \label{fig:phase_diagram}
\end{figure}

\begin{figure}[t]
    \centering
    \begin{tabular}{l}
    (a)  $s$-wave state\\
        \\
    \includegraphics[width=0.4\textwidth]{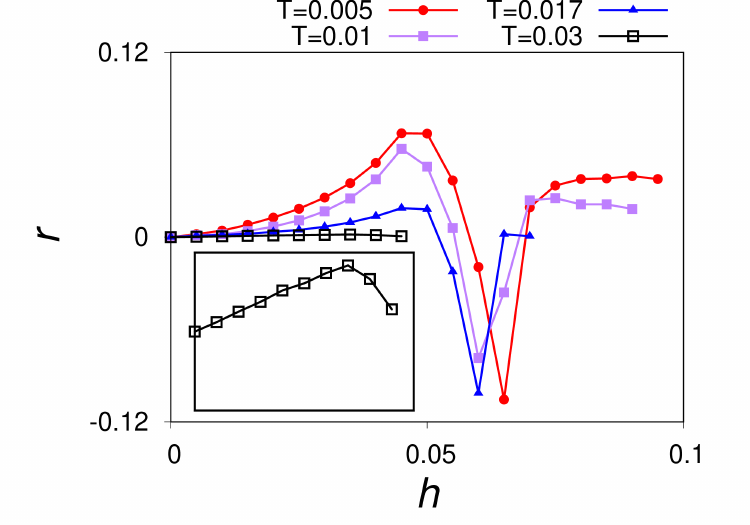}\\
        (b)  $d$-wave state\\%
        \\
        \includegraphics[width=0.4\textwidth]{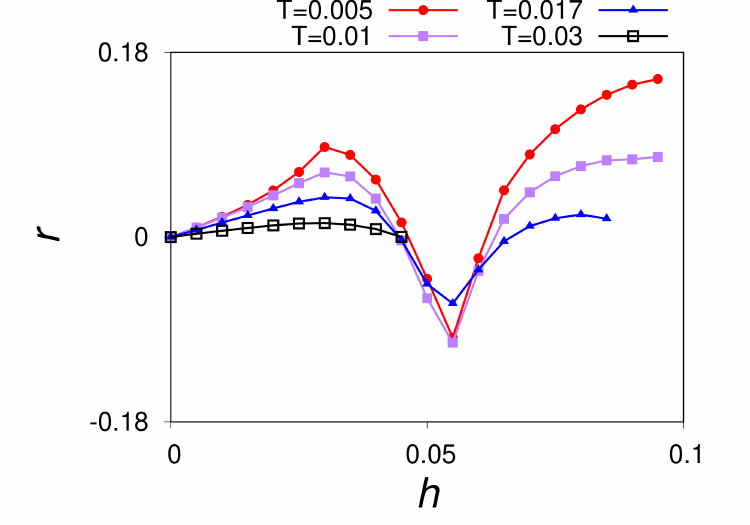}
    \end{tabular}
    \caption{The magnetic-field dependence of the quality factor $r$ for $T=0.005$ [red disk], $0.01$ [purple closed square], $0.017$ [blue closed triangle], and $0.03$ [black open square].
    Panels (a) and (b) correspond to the $s$-wave and $d$-wave states, respectively.
    The inset of the panel (a) indicates the results for $T=0.03$ shown in the domain $0\le h\le 0.05$ and $|r|\le0.002$.
    We adopted $L_y=200$ and $L_x=18000$ for the $s$-wave state at $T=0.03$, $L_x=12000$ for the $d$-wave state at $T=0.005$ and $0.01$, while $L_x=6000$
    otherwise.
    }
    \label{fig:r_of_h}
\end{figure}

We show in Fig.~\ref{fig:phase_diagram} the temperature and magnetic-field dependence of the diode quality factor $r$ [Eq.~\eqref{eq:def_r}]
and the equilibrium Cooper-pair momentum $q_0$ [Eq.~\eqref{eq:def_q0}]
for $s$-wave and $d$-wave superconducting states.
A finite value of $q_0$ indicates the realization of the helical superconductivity.
Figures~\ref{fig:phase_diagram}(a) and \ref{fig:phase_diagram}(b) are for the $s$-wave state, which reproduce the results of Ref.~\onlinecite{Daido2022-ox}.
Figures~\ref{fig:phase_diagram}(c) and \ref{fig:phase_diagram}(d) are for the $d$-wave state.
To see the quantitative details of the quality factor, we also show in Figs.~\ref{fig:r_of_h}(a) and \ref{fig:r_of_h}(b) the magnetic-field dependence of the quality factor at various temperatures for the $s$-wave and $d$-wave states, respectively.

Overall, the phase diagram for the $d$-wave state is qualitatively similar to the $s$-wave one, 
accompanying sign reversals {of the SDE} as increasing $h$.
The behavior of $q_0(T,h)$ is also similar to the $s$-wave case, and the nature of the helical superconductivity $q_0\neq 0$ is different between the low- and high-field states:
The value of $q_0$ grows rapidly for $h\gtrsim0.05$ in the $d$-wave state.
Following Ref.~\onlinecite{Daido2022-ox}, we call $h\sim0.05$ the ``crossover line" of helical superconductivity.
As discussed in the $s$-wave case~\cite{Daido2022-ox}, the crossover line is indeed a crossover around the transition temperature, while changes to the first-order transition line at low temperatures in our model.
The crossover region $T\gtrsim0.005$ is broader in the $d$-wave case than in the $s$-wave case ($T\gtrsim0.015$)~\cite{Daido2022-ox}.
{In Fig.~\ref{fig:phase_diagram}(c), the first sign reversal of $\Delta j_{\rm c}$ with increasing $h$ roughly coincides with the crossover line.}
Thus, the intrinsic SDE captures the precursor of the change in the helical superconducting states, regardless of the pairing symmetry.
In particular, the sign reversal occurs even around the transition temperature, indicating that the first-order transition is not a necessary ingredient.
Such a coincidence of the first sign reversal and the crossover line is also obtained in the quasiclassical calculation of the isotropic Rashba model, where the first-order transition is not reported~\cite{Ilic2022-kh}.
Therein, the second sign reversal of {the SDE} as increasing $h$ is absent, implying that the diode effect of the high-field helical state is sensitive to the details of the model and analysis.

The quantitative aspects of the SDE are different between the $s$- and $d$-wave superconductivity.
A larger nonreciprocity tends to be realized on the whole phase diagram in the $d$-wave state
as shown in Figs.~\ref{fig:phase_diagram}(a) and \ref{fig:phase_diagram}(c) as well as in Figs.~\ref{fig:r_of_h}(a) and \ref{fig:r_of_h}(b), which might be related to the presence of excitation nodes.
In particular, under low magnetic field near the transition temperature, the nonreciprocity is significantly larger in the $d$-wave state than that in the $s$-wave state.
This is understood by considering the vanishing $O(h)$ SDE in the ideally isotropic Rashba $s$-wave superconductor in the GL theory~\cite{Ilic2022-kh}.
While a finite $O(h)$ SDE around $T_{\rm c}$ is possible in this model due to the tetragonal anisotropy, it is expected that the $s$-wave $O(h)$ SDE still tends to be small: Indeed, the result for $T=0.03$ in Fig.~\ref{fig:r_of_h}(a) shows a $h$ linear diode effect with a tiny gradient.
The sizable diode effect of the $s$-wave state at lower temperatures seems to mainly follow from the $O(h^3)$ contributions.
On the other hand, the $d$-wave form factor avoids such an accidental cancellation of the $O(h)$ SDE, realizing a larger nonreciprocity even near the transition temperature [Fig.~\ref{fig:r_of_h}(b)].
Another difference {from the $s$-wave state} is the higher nonreciprocity in the low-$T$ and high-$h$ regime, indicating a large nonreciprocity of the high-field helical superconductivity.
This may also be related to the excitation nodes, while its precise reason remains to be clarified.

\begin{figure}[t]
    \centering
    \begin{tabular}{l}
    (a)  $s$-wave state\\
        \\
    \includegraphics[width=0.4\textwidth]{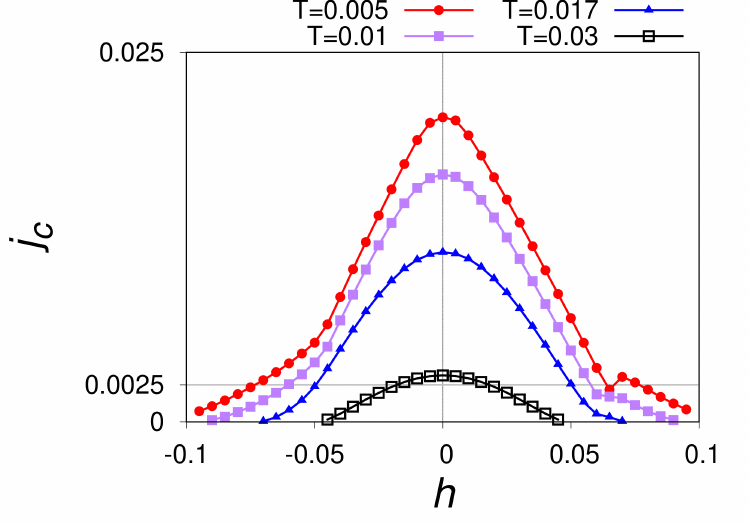}\\
        (b)  $d$-wave state\\%
        \\
        \includegraphics[width=0.4\textwidth]{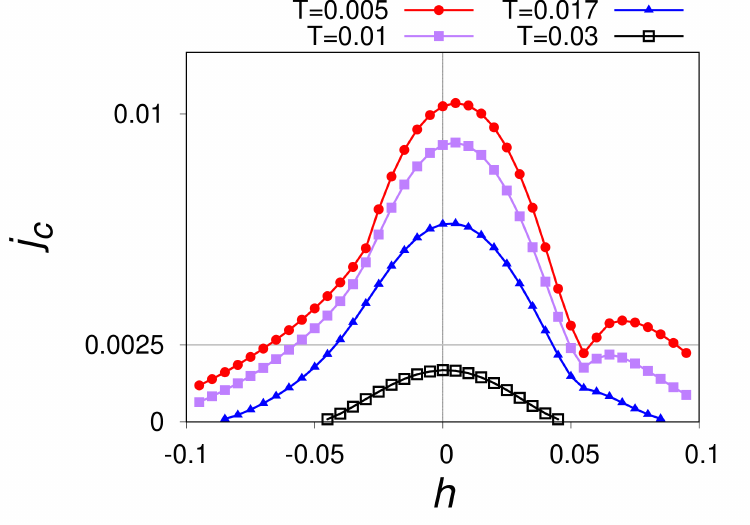}
\end{tabular}
    \caption{The magnetic-field dependence of the critical currents $j_{{\rm c} \pm}$ for the (a) $s$-wave and (b) $d$-wave states, which are shown in a single curve $j_{\rm c}(h)$ defined in Eq.~\eqref{eq:jc_both}.
    The notations and parameters are the same as those of Fig.~\ref{fig:r_of_h}.
    The value $j=0.0025$ is indicated by the gray horizontal lines for discussions in the text.
    }
    \label{fig:jp_jm_of_h}
\end{figure}

\subsection{Re-entrant superconductivity under supercurrent}
\label{subsec:re-entrantSC}
To further understand the results, we show in Figs.~\ref{fig:jp_jm_of_h}(a) and \ref{fig:jp_jm_of_h}(b) the critical currents in the $s$-wave and $d$-wave states, respectively.
The critical currents $j_{{\rm c} \pm}$ are shown in a single curve defined by
\begin{align}
    j_{c}(h)\equiv\begin{cases}
    j_{{\rm c} +}(h)&(h\ge0)\\
    -j_{{\rm c} -}(-h)&(h<0)
    \end{cases}.
    \label{eq:jc_both}
\end{align}
The low-field behavior of the quality factor discussed above is also clear in Fig.~\ref{fig:jp_jm_of_h}, where the $O(h)$ asymmetry of $j_{\rm c}(h)$, for example, is much more visible in the $d$-wave state than in the $s$-wave state.

An intriguing feature is the appearance of the second peak of $j_{{\rm c} +}$ under large magnetic fields, which are obtained for both the $s$- and $d$-wave states at $T=0.005$ [red points]: In other words, re-entrant superconductivity is realized in the $(j,h)$ phase diagram, since we have a stable superconducting solution for arbitrary parameters $(j,h)$ inside the $j_{\rm c}(h)$ curve.
The additional peak structure gets suppressed as increasing the temperature, and changes to an inflection point [see the result of $T=0.017$ shown by blue points].
Note that the sign reversal of the SDE occurs regardless of the presence or absence of the second peak, as is clear in Fig.~\ref{fig:r_of_h}.

We have indicated in Figs.~\ref{fig:jp_jm_of_h}(a) and \ref{fig:jp_jm_of_h}(b) the value of $j=0.0025$ by the horizontal gray lines.
As increasing the magnetic field at low temperatures with the fixed applied supercurrent $j=0.0025$, both the $s$-wave and $d$-wave superconducting solutions cease to exist once at $h\lesssim0.05$, and begin to exist again at $h\gtrsim 0.05$.
This implies that the re-entrant superconductivity is also realized in the $T-h$ phase diagram under the applied positive supercurrent, which makes clear contrast to the system under the negative supercurrent where a conventional transition line is expected {for $h \ge 0$. To the contrary, the re-entrant transition occurs only under the negative supercurrent for $h \le 0$.}
The drastic nonreciprocity of the transition lines under the supercurrent will be discussed in details in Sec.~\ref{subsec:critical_field_moderate}.

\begin{figure}
    \centering
    \includegraphics[width=0.45\textwidth]{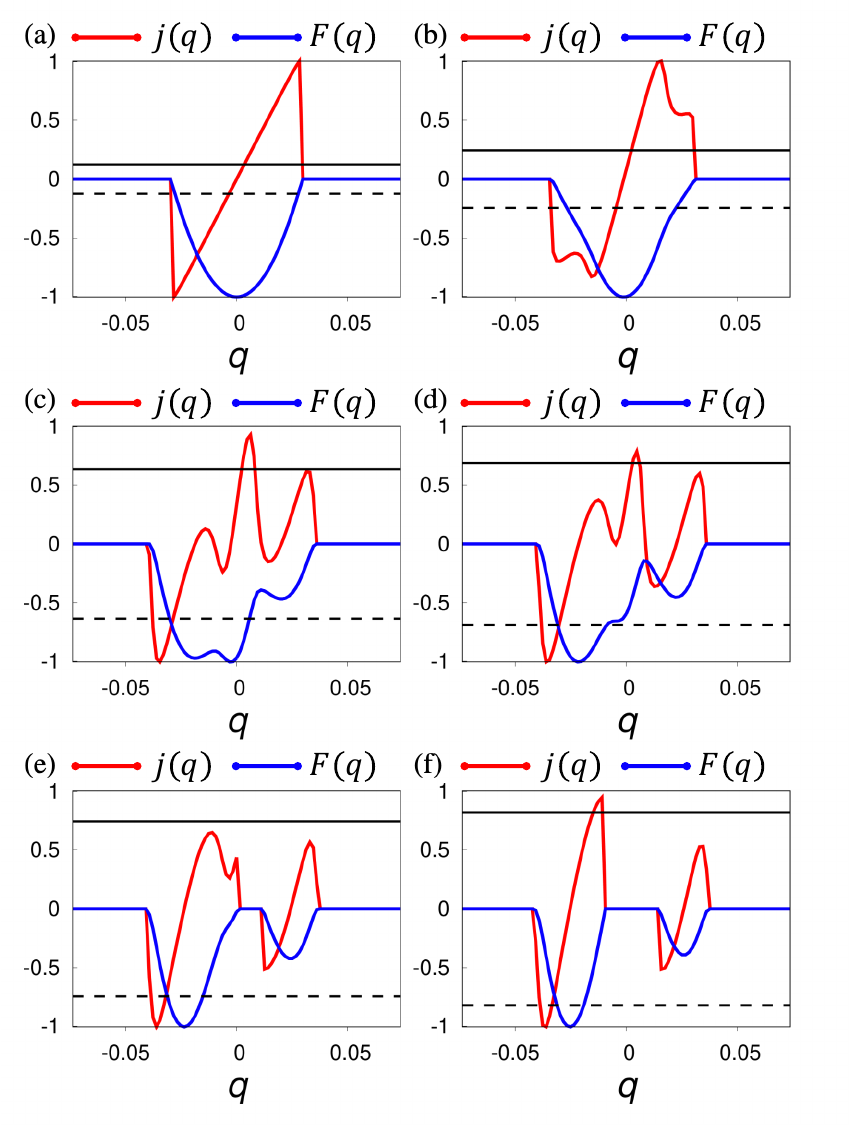}
    \caption{Momentum dependence of the supercurrent $j(q)$ [red color] and the condensation energy $F(q)$ [blue color] for the $s$-wave state at $T=0.005$.
    $j(q)$ and $F(q)$ are normalized by the maximum of their absolute value.
    The value of $\pm j$ with $j=0.0025$ are shown with black solid and dotted lines, respectively, in the same unit as $j(q)$.
    Calculations are done with 
    (a) $h=0$, (b) $h=0.04$, (c) $h=0.06$, (d) $h=0.0625$, (e) $h=0.065$, and (f) $h=0.0675$.
    }
    \label{fig:Fq_jq_s}
\end{figure}

To microscopically understand the re-entrant behavior of $j_{{\rm c} +}$ under high magnetic fields, 
we show in Fig.~\ref{fig:Fq_jq_s} the $q$ dependence of the electric current $j(q)$ and the condensation energy $F(q)$ of the $s$-wave states for various magnetic fields.
\begin{figure}
    \centering
    \includegraphics[width=0.45\textwidth]{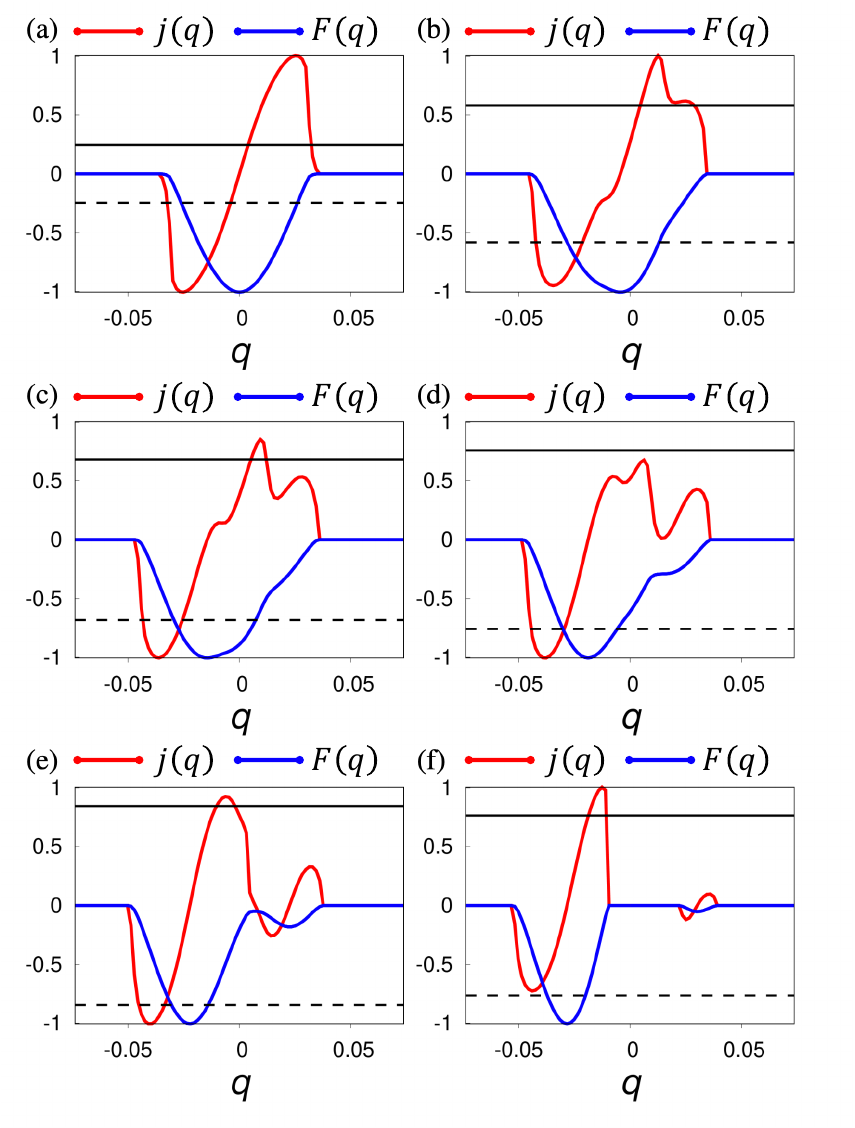}
    \caption{$j(q)$ and $F(q)$ for the $d$-wave state at $T=0.005$.
    Notations are the same as those of Fig.~\ref{fig:Fq_jq_s}.
    Calculations are done with (a) $h=0$, (b) $h=0.045$, (c) $h=0.05$, (d) $h=0.055$, (e) $h=0.06$, and (f) $h=0.07$.}
    \label{fig:Fq_jq_d}
\end{figure}
As demonstrated in Ref.~\onlinecite{Daido2022-ox}, the $q$ dependence of the condensation energy $F(q)$ shows a characteristic structure: Under low magnetic fields [panels (a) and (b)], $F(q)$ has a single-well structure as in the conventional BCS superconductivity, while consists of two wells under high magnetic fields [panels (e) and (f)].
As increasing the magnetic field, the single-well structure [panels (a) and (b)] changes to the triple-well structure [panels (c) and (d)], where the center well corresponds to the residue of the low-field helical state, and the left and right wells are the precursor of the high-field helical states.  
In agreement with the number of wells, around bottoms of which $F(q)$ is approximately quadratic, the function $j(q)=\partial_qF(q)$ consists of one, three, and two nearly straight lines and their interpolation in the low-, moderate-, and strong-magnetic-field regions, respectively.
Essentially the same structures are obtained for the $d$-wave state as well, where the multiple-well structure is less obvious due to the smearing by the nodal excitation of quasiparticles [Fig.~\ref{fig:Fq_jq_d}].

In Figs.~\ref{fig:Fq_jq_s} and \ref{fig:Fq_jq_d}, we indicate by the black solid and dotted horizontal lines the value of supercurrent $\pm j=\pm0.0025$.
It is clear that the system ceases to have a solution of the positive supercurrent $j(q)=+j$ as increasing $h$, before begins to have a solution again under higher magnetic fields.
This gives rise to the second peak of $j_{{\rm c} +}$ , or the re-entrant behavior  of the superconducting states, which are seen in Fig.~\ref{fig:jp_jm_of_h}.
By contrast, all the panels show a superconducting solution for the negative current direction, $j(q)=-j$, and therefore no re-entrant structure appears.

The essential difference of the positive and negative current directions is understood as follows.
Let us assign the local maximum $\mathcal{J}_{\mathrm{max}}$ and local minimum $\mathcal{J}_{\mathrm{min}}$ of $j(q)$ to each well of $F(q)$.
With the labels $l,c,r$ specifying the left, center, and right wells, we obtain the critical current $j_{{\rm c} +}$, i.e. the global maximum of $j(q)$, by
\begin{align}
    j_{{\rm c} +}
    =\mathcal{J}_{\mathrm{max}}(c),
\end{align}
for Figs.~\ref{fig:Fq_jq_s}(a)-(d), and
\begin{align}
    j_{{\rm c} +}
    =\mathcal{J}_{\mathrm{max}}(l),
\end{align}
for Figs.~\ref{fig:Fq_jq_s}(e) and \ref{fig:Fq_jq_s}(f).
In the same way, 
we also obtain
\begin{align}
    j_{{\rm c} -}
    =\mathcal{J}_{\mathrm{min}}(c),
\end{align}
for Figs.~\ref{fig:Fq_jq_s}(a) and \ref{fig:Fq_jq_s}(b), while
\begin{align}
    j_{{\rm c} -}
    =\mathcal{J}_{\mathrm{min}}(l),
\end{align}
for Figs.~\ref{fig:Fq_jq_s}(c)-(f).
The important point is that $\mathcal{J}_{\mathrm{max}}(l)$ is prevented from developing due to the presence of the central well, as shown in Figs.~\ref{fig:Fq_jq_s}(c)-(e), while $\mathcal{J}_{\mathrm{min}}(l)$ is not.
As $h$ is increased, the central well becomes destabilized and $\mathcal{J}_{\mathrm{max}}(c)$ becomes smaller.
At the same time, $\mathcal{J}_{\mathrm{max}}(l)$ is allowed to develop.
Thus, the characteristic Cooper-pair-momentum dependence of the free energy accompanied by the change in the helical state is the origin of the re-entrant critical current in the positive direction.
On the other hand, there is no such complexity
in the development of $\mathcal{J}_{\mathrm{min}}(l)$,
and thus the re-entrant behavior {in the critical current} is absent.
The $d$-wave case can also be understood in the same way, as is clear in Fig.~\ref{fig:Fq_jq_d}.

\subsection{First-order transition and crossover in the superconducting state under supercurrent}
\label{subsec:re-entrantSC2}
\begin{figure}
    \centering
    \begin{tabular}{l}
        (a)\\
            \includegraphics[width=0.4\textwidth]{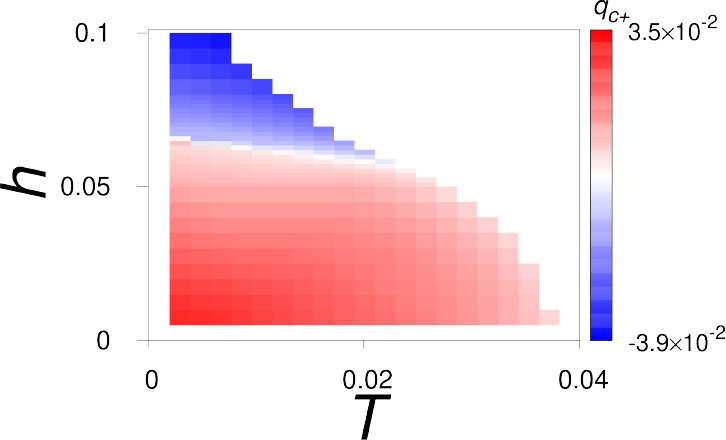}\\
            (b)\\
    \includegraphics[width=0.4\textwidth]{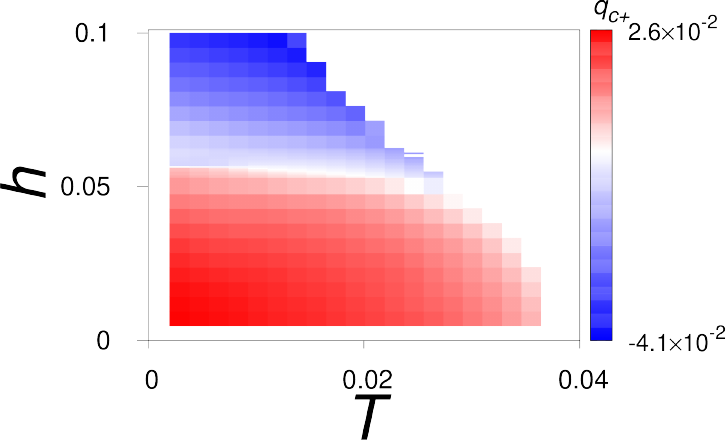}
        \end{tabular}
    \caption{$T$ and $h$ dependence of the critical momentum $q_{{\rm c}+}$ for (a) $s$-wave and (b) $d$-wave state.
    The mesh of $h$ is taken dense near the crossover line of $q_0$.
    }
    \label{fig:phase_diagram_q+}
\end{figure}

Interestingly, as discussed above, the superconducting solution supporting the critical current $j_{{\rm c} \pm}$ changes from the central to the left well as $h$ is increased (see Figs.~\ref{fig:Fq_jq_s} and~\ref{fig:Fq_jq_d}).
We show in Fig.~\ref{fig:phase_diagram_q+} the phase diagram for the critical momenta $q_{{\rm c}+}$ defined by
\begin{align}
    j(q_{{\rm c}+})\equiv j_{{\rm c} +}.
\end{align}
The positive (negative) values of $q_{{\rm c}+}$ indicate that the critical current $j_{{\rm c} +}$ is realized in the superconducting solution related to the low-field (high-field) helical state, i.e. the central (left) well of $F(q)$.
At low temperatures, $T\lesssim0.015$ for the $s$-wave state [$T\lesssim0.01$ for the $d$-wave state], the sign change of
$q_{{\rm c}+}$ occurs abruptly at $h\sim0.064$ [$h\sim0.057$] as increasing the magnetic field.
Figures~\ref{fig:phase_diagram_q+}(a) and \ref{fig:phase_diagram_q+}(b) show that such a change 
in $q_{{\rm c}+}$ becomes broad at higher temperatures.
This can be understood from Figs.~\ref{fig_app:Fq_jq_s} and \ref{fig_app:Fq_jq_d} in Appendix~\ref{app:fig_high_T} showing that the three wells consisting of condensation energy $F(q)$ at low temperatures are merged into a single well at higher temperatures probably due to the thermal quasiparticles.
Even in such a situation, the sign of $q_{{\rm c}+}$ can still be used as the rough standard to judge the nature of the superconducting solution supporting the critical current $j_{{\rm c} +}$.
It is also found that $q_{{\rm c}-}$ defined by $j(q_{{\rm c}-})\equiv j_{{\rm c} -}$
shows an abrupt change at low temperatures, which occurs at lower magnetic field than that of $q_{{\rm c}+}$: $T\lesssim0.015$ with $h\sim0.047$ [$T\lesssim0.01$ with $h\sim0.03$].
\begin{figure*}
    \centering  
    \includegraphics[width=0.9\textwidth]{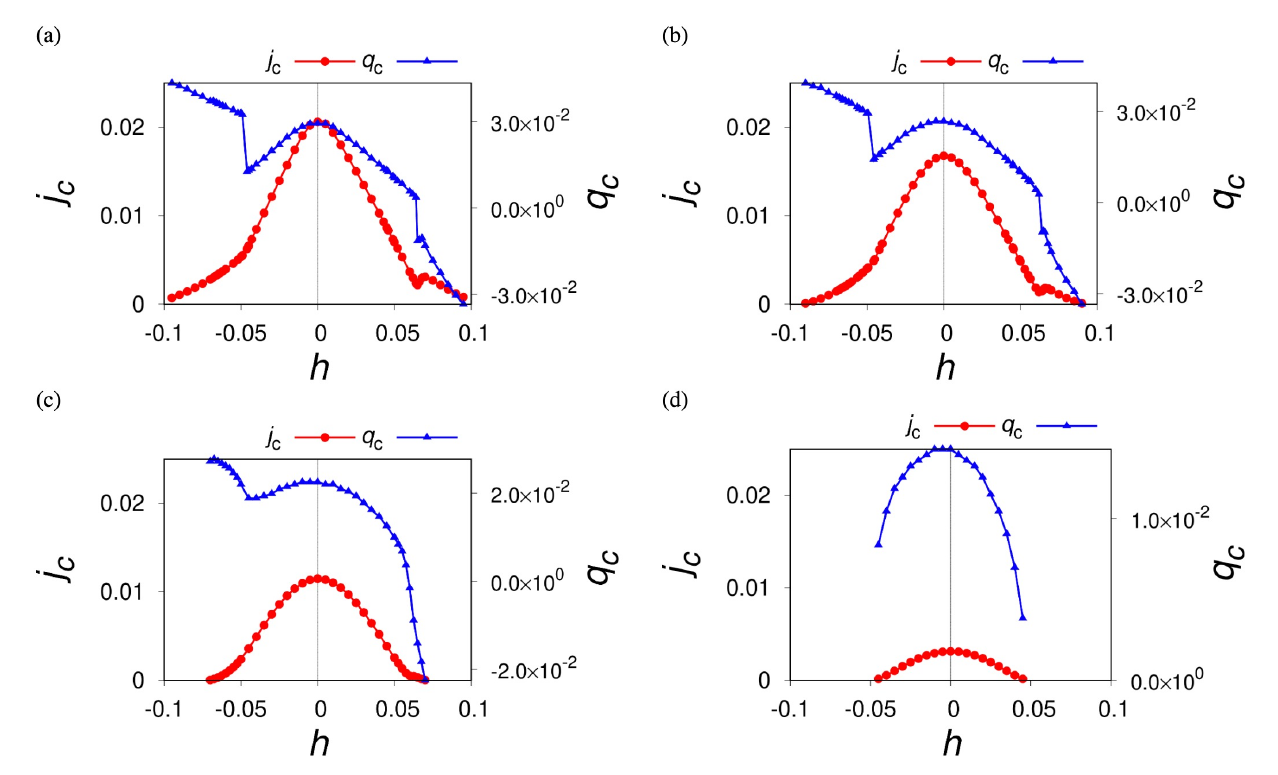}
    \caption{$h$ dependence of the critical current $j_{\rm c}(h)$ [red disks] and $q_{\rm c}(h)$ [blue triangles] in the $s$-wave state at (a) $T=0.005$, (b) $T=0.01$, (c) $T=0.017$, and (d) $T=0.03$.
    The solid lines are the guide for the eye.
    }
    \label{fig:jc_qc_s}
\end{figure*}
\begin{figure*}
    \centering
    \includegraphics[width=0.9\textwidth]{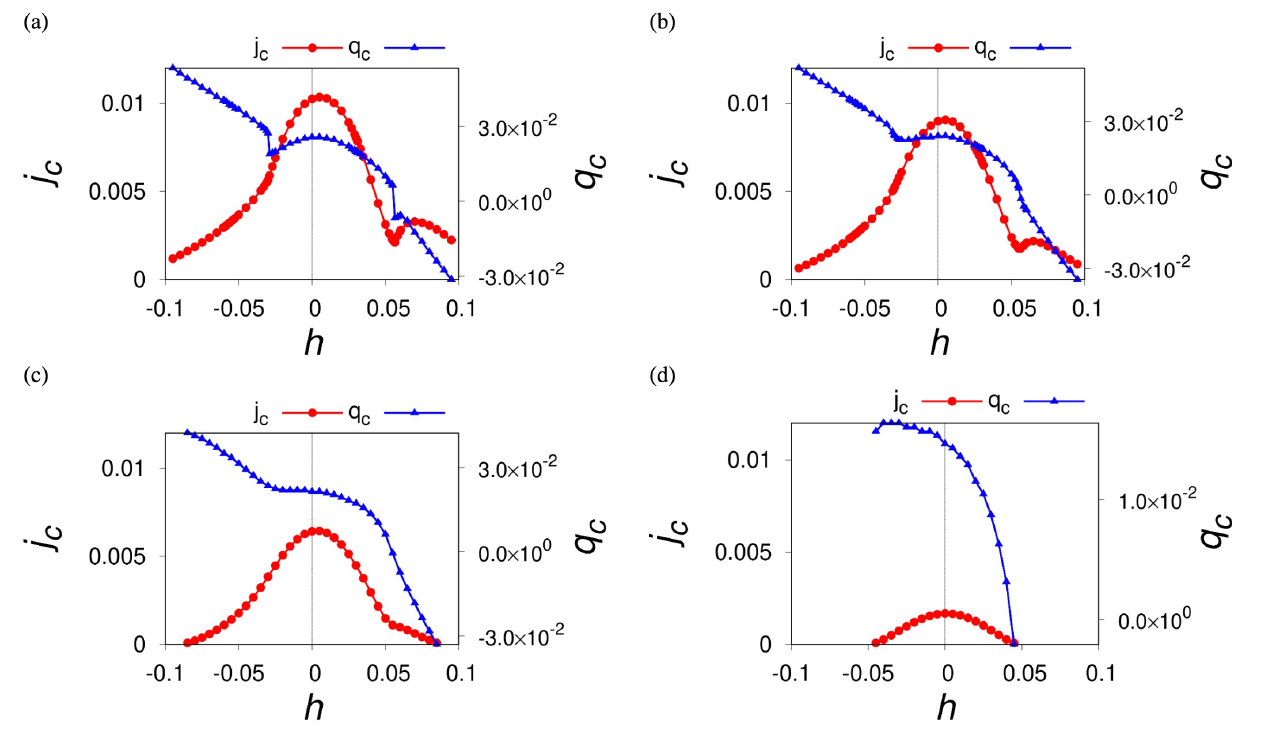}
    \caption{$h$ dependence of the critical current $j_{\rm c}(h)$ and $q_{\rm c}(h)$ in the $d$-wave state at (a) $T=0.005$, (b) $T=0.01$, (c) $T=0.017$, and (d) $T=0.03$.
    Notations are the same as those in Fig.~\ref{fig:jc_qc_s}.
    }
    \label{fig:jc_qc_d}
\end{figure*}

The behavior of $q_{c\pm}$ is faithfully reflected in that of $j_{{\rm c} \pm}(h)$.
We show in Figs.~\ref{fig:jc_qc_s} and \ref{fig:jc_qc_d} the $h$ dependence of $j_{c}(h)$ defined in Eq.~\eqref{eq:jc_both} and $q_{\rm c}(h)$ defined by
\begin{align}
q_{\rm c}(h)\equiv\begin{cases}
    q_{{\rm c}+}(h)&(h\ge0)\\
    -q_{{\rm c}-}(-h)&(h<0)
    \end{cases},
\label{eq:q_c_both}    
\end{align}
for various temperatures.
Figures~\ref{fig:jc_qc_s}(a) and \ref{fig:jc_qc_d}(a) show the results for the $s$-wave and $d$-wave states at a low temperature $T=0.005$, respectively.
The onset of the second peak of $j_{{\rm c} +}$ coincides with the jump of $q_{{\rm c}+}$.
This means that there occurs a first-order phase transition when we change the parameters $(j,h)$ near the critical current $j_{\rm c}(h)$.
In other words, the first and second domes of $j_{\rm c}$ actually belong to the different superconducting states at low temperatures, namely the low- and high-field helical superconductivity.
{Similarly, first-order transition related to $q_{{\rm c}-}$ is also obtained near $h\sim0.05$ [$h\sim0.03$], which results in the small jump of $dj_{{\rm c} -}(h)/dh$.} 

Figures~\ref{fig:jc_qc_s}(b) and \ref{fig:jc_qc_s}(c) and Figs.~\ref{fig:jc_qc_d}(b) and  \ref{fig:jc_qc_d}(c) show $j_{\rm c}(h)$ and $q_{\rm c}(h)$ for higher temperatures $T=0.01$ and $T=0.017$ for the $s$- and $d$-wave states.
As increasing the temperature, the $q_{\rm c}(h)$ curve becomes smooth, and therefore the first-order transition obtained for $T=0.005$ changes to a rapid crossover.
The crossover of $q_{c\pm}(h)$ is reflected into $j_{\rm c}(h)$ as its inflection points.
By further increasing the temperature, at $T=0.03$, the system does not experience a crossover of $q_{{\rm c}+}$ as is clear in Fig.~\ref{fig:phase_diagram_q+}, and thus the critical current $j_{\rm c}(h)$ shows a {conventional magnetic-field dependence.}

In summary, the sign reversal of the SDE and the re-entrant superconductivity under the supercurrent are caused by the development of the characteristic structure in the $q$-dependence of the condensation energy.
They are the results of the crossover and the first-order transition of the critical momentum, {which are qualitatively equivalent to those of $q_0$,} namely the change of the helical superconductivity.
Observation of these phenomena would give a strong evidence of the helical superconductivity, {which has been awaited for a long time.}
The current-induced (or -enforced) first-order transitions between the low- and high-field helical states would be directly detectable by observation of the anomaly in, e.g., the optical conductivity.
We leave the detailed study of the current-induced transitions to the future issue.

\section{Nonreciprocal transition lines under supercurrent}
\label{sec:critical_field}
Based on the understanding of the depairing critical current in the previous section,
we discuss the nonreciprocity of the temperature-magnetic-field phase diagram
under the applied supercurrent.
A critical magnetic field $h_{{\rm c}+}(T)$ under the current $+|j|$ can be determined by using the temperature and magnetic-field dependence of the critical current $j_{{\rm c} +}(T,h)$.
Actually, by increasing $h$ with fixing $T$, $j_{{\rm c} +}(T,h)$ becomes smaller than the given value of $+|j|$ at a value of $h$, above which no superconducting states can support the electric current $+|j|$.
Thus, we define
\begin{align}
    j_{{\rm c} +}(T,h_{{\rm c}+}(T))=+|j|.\label{eq:Hc2p}
\end{align}
We also define
\begin{align}
j_{{\rm c} -}(T,h_{{\rm c}-}(T))=-|j|,\label{eq:Hc2m}
\end{align}
for the critical magnetic field $h_{{\rm c}-}(T)$ under the current $-|j|$.
Similarly, the transition temperature $T_{\rm c}^\pm(H)$ under the current $\pm|j|$ is obtained by
\begin{align}
    j_{{\rm c} +}(T_{{\rm c}+}(h),h)&=+|j|,\label{eq:Tcp}\\
j_{{\rm c} -}(T_{{\rm c}-}(h),h)&=-|j|.\label{eq:Tcm}
\end{align}
Accordingly, the transition line in the $T-h$ phase diagram is determined by the series of points
\begin{align}
    &\set{(T,h)=(T,h_{c\pm}(T))}
    =\set{(T,h)=(T_{c\pm}(h),h)}.
\end{align}
In the following, we adopt Eqs.~\eqref{eq:Hc2p}-\eqref{eq:Tcm} as the definition of the critical magnetic fields and the transition temperatures under the supercurrent, and discuss their relation with the SDE.

Note that the transition lines under supercurrent defined here are the contour lines of the nonreciprocal critical currents $j_{{\rm c} \pm}(T,h)$.
Thus, they can be experimentally obtained either by the critical current measurements for various $T$ and $h$, or by directly determining the $T-h$ phase diagram under a fixed applied supercurrent. 
{Experiments on $H_{c2}$ under the supercurrent have actually been performed in Refs.~\onlinecite{Miyasaka2021-ly,Kawarazaki_submitted}.
The transition line in the $(h,j)$ space coincides with that obtained by the critical-current experiments under the magnetic field~\cite{Kawarazaki_submitted}, supporting the validity of Eqs.~\eqref{eq:Hc2p}-\eqref{eq:Tcm}.
Note also that similar techniques have recently been used to measure the anisotropy of $H_{c2}$ in Sr${}_2$RuO${}_4$ under the electric current~\cite{Araki2021-lk}.
}

It should be noted that the transition lines refer to two nonequivalent functions $j_{{\rm c} \pm}(T,h)$, and thus in principle have more information than that can be read out only from $\Delta j_{\rm c}(T,h)$ or $r(T,h)$.
In the following, it turns out that there is a one-to-one correspondence between nonreciprocal transition lines and the SDE when the diode quality factor $r(T,h)$ is small, where a simple purterbative calculation allows us an phenomenological understanding.
On the other hand, it may not be the case when $r(T,h)$ becomes large, and the nonreciprocal transition lines can deeply reflect the microscopic nature of the superconducting state.

For the case of the intrinsic SDE of helical superconductivity, the behavior of the nonreciprocal transition lines can be classified into three regimes determined by the strength of the electric current.
The first is the large $|j|$ regime, where the superconducting state is stable only under low temperature and low magnetic field.
In this case, the transition line is off the crossover regime of the helical superconductivity, and no anomalous behavior is expected.
The second case is the small $|j|$ regime.
In this case, the transition line is almost the same as that in the absence of the current, and thus
less anomalous behavior is expected.
The third case is the regime with intermediate strength of $|j|$, where the transition line deeply crosses the crossover line of the helical superconductivity.
In this region, the characteristic momentum dependence of the free energy is developed, and therefore an anomalous behavior of the transition lines is expected.
In the following, we first show the results of the transition lines under small and large electric currents and
discuss the phenomenology of the nonreciprocal transition lines.
Then we show the results for the intermediate strength of the supercurrent, and demonstrate the behavior characteristic of helical superconductivity.

\subsection{Small and large supercurrent: Phenomenological theory of the nonreciprocal transition lines}
\label{subsec:phenomenology_critical_field}

\begin{figure}
    \centering
    \begin{tabular}{l}
    (a)\\
        \includegraphics[width=0.45\textwidth]{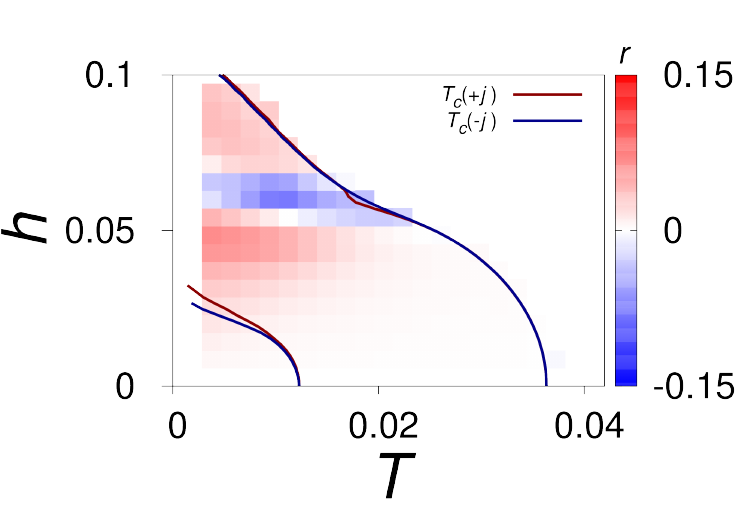}\\
        (b)\\
    \includegraphics[width=0.45\textwidth]{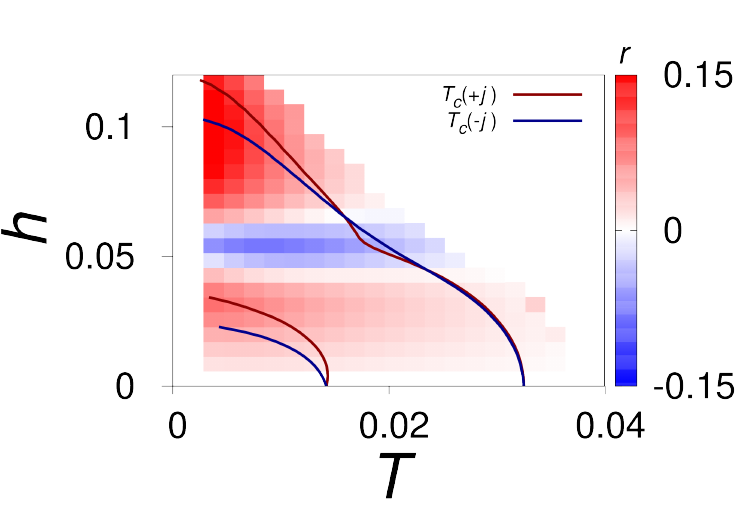}
        \end{tabular}
    \caption{Nonreciprocal transition lines of the (a) $s$-wave and (b) $d$-wave states under small and large electric currents.
    The dark-red and dark-blue curves show the transition lines with the positive and negative electric currents, respectively, and are shown on the phase diagram in Figs.~\ref{fig:phase_diagram}(a) and \ref{fig:phase_diagram}(c) for comparison with the diode quality factor $r(T,h)$.
    The outer and inner lines correspond to (a) $j=0.0005$ and $j=0.015$; (b) $j=0.001$ and $j=0.0075$, respectively.
    }
    \label{fig:transition_line_dx2y2_SL}
\end{figure}

We show in Figs.~\ref{fig:transition_line_dx2y2_SL}(a) and \ref{fig:transition_line_dx2y2_SL}(b) the transition lines under small and large electric currents for the $s$- and $d$-wave states, where the dark-red and dark-blue lines correspond to the positive and negative current directions, respectively.
The inner (outer) lines correspond to the result for the large (small) electric current.
The transition lines are drawn on the phase diagrams of the diode quality factor $r(T,h)$, Figs.~\ref{fig:phase_diagram}(a) and \ref{fig:phase_diagram}(c).
A large split of the transition lines
\begin{align}
\delta h_{{\rm c}}(T)\equiv h_{{\rm c}+}(T)-h_{{\rm c}-}(T),
\end{align}
is obtained where the diode quality factor $r(T,H)$ is large in the phase diagram.
We also observe the crossings of the transition lines near the crossover line.
It is also interesting to note that the superconducting state under the positive current direction [dark red curve]
{shows an enhancement of transition temperature} under the small finite magnetic field, 
which is evident for the $d$-wave state under a large supercurrent [the inner dark-red transition line of Fig.~\ref{fig:transition_line_dx2y2_SL}(b)]. 
{This indicates the field-enhanced superconductivity.}

These behaviors follow from the general considerations based on the finite diode effect $r(T,H)\neq0$ and the definitions {of nonreciprocal transition lines} Eqs.~\eqref{eq:Hc2p}-\eqref{eq:Tcm}.
Let us first consider the crossing points of the nonreciprocal critical fields.
We define the averaged critical magnetic field by
\begin{align}
    \bar{h}_{{\rm c}}(T)&=\frac{1}{2}[h_{{\rm c}+}(T)+h_{{\rm c}-}(T)].
\end{align}
A crossing point at $T=T^*$ satisfies
\begin{align}
    h_{c\pm}(T^*)=\bar{h}_{{\rm c}}(T^*),\quad \delta h_{{\rm c}}(T^*)=0,
\end{align}
and thus,
\begin{align}
j_{{\rm c} \pm}(T^*,\bar{h}_{{\rm c}}(T^*))=\pm|j|.
\end{align}
Summing up $j_{{\rm c} +}$ and $j_{{\rm c} -}$, we obtain 
$\Delta j_{c}(T^*,h^*)=0$,
and
\begin{align}
r(T^*,h^*)=0,\label{eq:crossing_r}
\end{align}
for the crossing point
\begin{align}
    (T^*,h^*)=(T^*,h_{c\pm}(T^*))=(T_{c\pm}(h^*),h^*).
\end{align}
This indicates that the crossing points are generally located on the sign-reversal lines of the SDE.

In addition to the non-perturbative relation Eq.~\eqref{eq:crossing_r}, we can estimate the split of the critical fields $\delta h_{{\rm c}}(T)$ from the SDE,
assuming that $\delta h_{{\rm c}}(T)$ is sufficiently small.
By using the expansion
\begin{align}
j_{{\rm c} \pm}(h_{c\pm})=j_{{\rm c} \pm}(\bar{h}_{{\rm c}})\pm\frac{1}{2}\partial_hj_{{\rm c} \pm}(\bar{h}_{{\rm c}})\delta h_{{\rm c}}+O(\delta h_c)^2,
\end{align}
and the similar one for the critical temperature,
we obtain
\begin{align}
    \delta h_{{\rm c}}(T)&=\frac{r(T,\bar{h}_{{\rm c}}(T))}{-\partial_h\ln\bar{j}_{\rm c}(T,\bar{h}_{{\rm c}}(T))/2},\\
    \delta T_{c}(h)&=\frac{r(\bar{T}_{{\rm c}}(h),h)}{-\partial_T\ln\bar{j}_{\rm c}(\bar{T}_{{\rm c}}(h),h)/2},\label{eq:delta_Tc_r}
\end{align}
with the split and the average of the transition temperature $\delta T_{\rm c}(h)=T_{{\rm c}+}(h)-T_{{\rm c}-}(h)$ and $\bar{T}_{{\rm c}}(h)=(T_{\rm c}^+(h)+T_{\rm c}^-(h))/2$.
Thus, the split of the transition lines is determined by the SDE up to the first order.
In particular, their sign is the same as that of SDE, since the denominators are usually positive.
On the other hand, the above expansion goes worse when $r(T,\bar{H}_{c2})$ is of order unity and $\delta h_c(T),\delta T_{\rm c}(T)$ get large, where the simple coincidence between the SDE and nonreciprocal transition lines might break down, {as we show later.}

Next, we discuss the 
{field-enhanced superconductivity} by a small magnetic field which is seen for the $d$-wave state under the large supercurrent [the inner dark-red transition line of Fig.~\ref{fig:transition_line_dx2y2_SL}(b)].
Note that the relation
\begin{align}
T_{{\rm c}+}(h)=T_{{\rm c}-}(-h),\label{eq:Tcpm}
\end{align}
holds by the time-reversal symmetry, while a similar relation also holds for the critical magnetic field [the value of $|h_{{\rm c}+}|$ is equivalent to $|h_{{\rm c}-}|$ in the opposite field direction].
Equation~\eqref{eq:Tcpm} means that $T_{{\rm c}+}(h)$ for negative $h$ is obtained by $T_{{\rm c}-}(|h|)$, and thus a smooth extended transition line $T_{{\rm c}+}(h)$ is obtained when the dark-blue curve is flipped to the negative-$h$ region and connected with the dark-red curve.
It also follows from Eq.~\eqref{eq:Tcpm} that
$\bar{T}_{{\rm c}}(h)$ is an even function of $h$ and therefore peaked at $h=0$.
Note that $\delta T_{\rm c}(h)$ has the $h$-linear component as a result of {the $O(h)$ SDE} [see Eq.~\eqref{eq:delta_Tc_r}].
Then, the peak of {$T_{c\pm}(h)=\bar{T}_{{\rm c}}(h)\pm\delta T_{\rm c}(h)/2$} is shifted to a finite $h$, leading to the skewed transition line as in Fig.~\ref{fig:transition_line_dx2y2_SL}(b).
In our model, the $O(h)$ SDE is small in the $s$-wave state as well as the $d$-wave state near the transition temperature.
This is why the skewness is the most visible for the transition line which passes the low-temperature region of the phase diagram in the $d$-wave state.

{We would like to emphasize that} the behaviors of the nonreciprocal transition lines discussed in this section are understood only from the definitions~\eqref{eq:Hc2p}-\eqref{eq:Tcm}.
Thus, they
{are universal behaviors independent} of the origins of the diode effect.

\subsection{Moderate strength of the supercurrent: Nonreciprocal re-entrant transition lines
}
\label{subsec:critical_field_moderate}

\begin{figure}
    \centering
    \begin{tabular}{l}
    (a) $s$-wave state\\
        \includegraphics[width=0.45\textwidth]{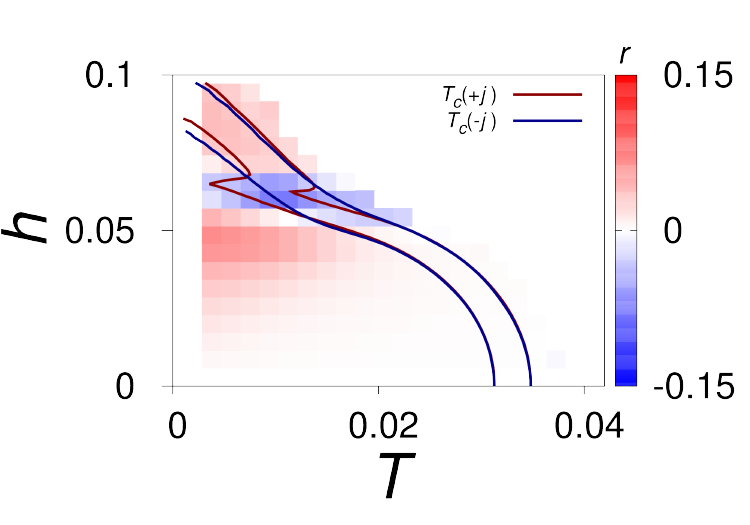}\\
        (b) $d$-wave state\\
        \includegraphics[width=0.45\textwidth]{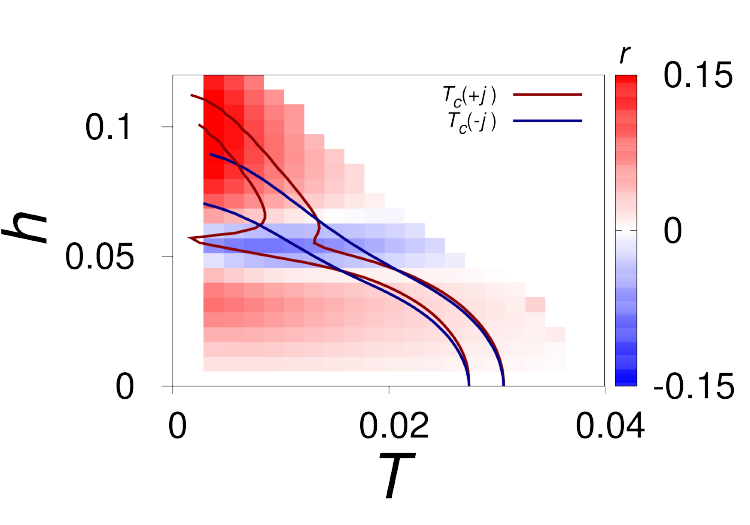}\\
    \end{tabular}
    \caption{Nonreciprocal transition lines under moderate strength of the electric currents for (a) the $s$-wave and (b) $d$-wave states.
    The notations are the same as those in Fig.\ref{fig:transition_line_dx2y2_SL}.
    The strength of the electric current (a) $j=0.001$ (outer lines) and $j=0.0025$ (inner lines) are adopted for the $s$-wave state while (b) $j=0.0015$ (outer lines) and $j=0.0025$ (inner lines) for the $d$-wave state.}
    \label{fig:transition_line2}
\end{figure}
\begin{figure}
    \centering
    \begin{tabular}{l}
    (a) $s$-wave state\\
        \includegraphics[width=0.45\textwidth]{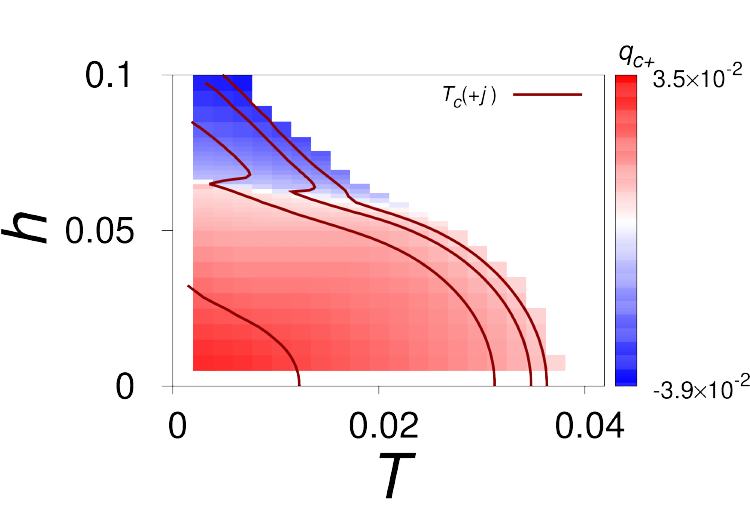}\\(b) $d$-wave state\\
    \includegraphics[width=0.45\textwidth]{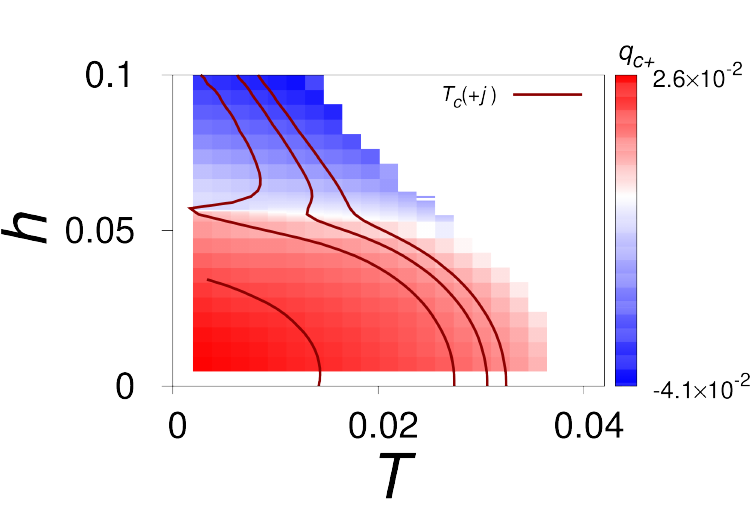}
    \end{tabular}
    \caption{The transition lines $T_{{\rm c}+}(h)$ for various supercurrent in the (a) $s$-wave and (b) $d$-wave states, shown on Figs.~\ref{fig:phase_diagram_q+}(a) and \ref{fig:phase_diagram_q+}(b), respectively.
    The electric current is set as (a) $j=0.0005$, {$0.001$}, $0.0025$, $0.015$ and (b) $j=0.001$, $0.0015$, $0.0025$, $0.0075$ from outside to inside of the phase diagram [same as Figs.~\ref{fig:transition_line_dx2y2_SL} and \ref{fig:transition_line2}].}
    \label{fig:jcp_qcp}
\end{figure}

In contrast to the small and large current regime {discussed in the previous subsection,}
the simple correspondence between the quality factor $r$ and the nonreciprocal transition lines generally breaks down for the moderate strength of the supercurrent.
In such a situation, it is expected that the nature of the superconducting state is deeply reflected into the behavior of the transition lines.
We show in Fig.~\ref{fig:transition_line2} the nonreciprocal transition lines under moderate electric currents for the (a) $s$-wave and (b) $d$-wave states.
The crossings of the transition lines are obtained on the sign-reversal lines of the diode effect, in accordance with Eq.~\eqref{eq:crossing_r}.
Interestingly, the transition lines for the positive currents [dark-red lines] show re-entrant behavior around the crossover line, in contrast to those for the negative currents [dark-blue lines].
Such current-induced re-entrant superconductivity is a direct consequence of the non-monotonic behavior of the $h$ dependence of $j_{{\rm c} +}$ [Fig.~\ref{fig:jp_jm_of_h}].

The low- and high-field superconducting domes correspond to the low- and high-field helical superconductivity, which
are connected by either a first-order transition or a crossover.
To see this, we
show in Figs.~\ref{fig:jcp_qcp}(a) and \ref{fig:jcp_qcp}(b) the transition line $T_{{\rm c}+}(h)$ in the $s$-wave and $d$-wave states, respectively,
on the phase diagram for the critical momentum $q_{{\rm c}+}$ [Fig.~\ref{fig:phase_diagram_q+}].
Clearly, the kinks of the transition lines are located on the sign-reversal line of $q_{{\rm c}+}$, both in the presence and absence of the re-entrant behavior.
This clearly indicates that the superconducting solution supporting the critical current $j_{{\rm c} +}$ changes from the low-field to high-field helical states when the magnetic field is increased along the transition line, leading to the anomaly of the transition line.
{On the $q_{{\rm c}+}=0$ line, the critical momentum $q_{{\rm c}+}$ jumps from a negative to positive value at low temperatures, while smoothly changes at higher temperatures.}
Thus, the kinks of the transition lines signal the crossover at smaller current and the first-order transition at larger current.
The kinks and re-entrant behavior of the transition lines are the strong evidence of the crossover of the helical superconductivity.

\section{Discussion}
\label{sec:discussion}
We have microscopically revealed the SDE and nonreciprocal transition lines in the Rashba-Zeeman model for the $s$-wave and $d$-wave superconductivity.
In this section, we comment on some technical aspects and future issues.

We first make some remarks on the effects neglected in our treatment of the SDE, which includes (1) the possibility of the multiple $q$ states as well as (2) the impurity effects.

(1) The present analysis assumes the Fulde-Ferrell (FF)-type pair potential of the form $\Delta(x)=e^{iqx}\Delta$.
In general, the pair potential may include several Fourier components as is known for the Larkin-Ovchinnikov (LO) state $\Delta(x)=\Delta \cos qx$ in centrosymmetric superconductors.
Indeed, study of the Rashba-Zeeman model by the quasiclassical approximation points to a stabilization of the LO-like state in a dome-shaped region near the crossover line at low temperatures~\cite{Agterberg2007-vl}.
The LO-like state is also studied in Refs.~\onlinecite{Dimitrova2003-mo,Dimitrova2007-hp} assuming density of states equivalent on the split Fermi surfaces (such a simplification leads to the uniform state with $q_0=0$ under low fields).
What we can say for sure is that the intrinsic SDE studied in this paper remains valid under low and high magnetic fields as well as near the transition temperature $H_{c2}(T)$, where the LO-like state is not stabilized.
Therefore, for example, the sign reversal around $H_{c2}(T)$ by the crossover of the helical superconducting state is unaffected, even when the multiple $q$ states are taken into account.

In principle, the SDE may be affected where the LO-like state is stabilized, and comprehensive study of its impact is left as a future issue.
Nevertheless, 
qualitative results may not be changed by the inclusion of such multiple-$q$ degrees of freedom to the gap equation.
Note that the near degeneracy of the (quasi-stable) single-$q$ solutions with the momentum $\sim\pm q_0$ is important for the stabilization of the LO-like state:
Indeed, the difference of the density of states on the Rashba-split Fermi surfaces must be small to stabilize the LO-like solution~\cite{Agterberg2007-vl}.
Application of the electric current will lift such a near degeneracy, and
thus the order parameter more close to the FF-type one would determine the depairing critical current.
The treatment of this paper might give a good approximation even if small admixing of the other Fourier components is present, and
hence, the phase diagram for $\Delta j_{\rm c}$ and nonreciprocal transition lines {are} 
less affected than the thermodynamic phase diagram in the absence of the electric current.
It is also an interesting future issue to theoretically/experimentally identify the current-induced transition from the LO-like state to the single-$q$ helical superconducting state. 

(2) The present analysis focuses on the SDE in the clean limit. 
An important future direction is the effect of disorders and impurities on the SDE.
It is known that the high-field helical superconductivity is fragile against impurities, while the low-field state is robust against moderate disorders, i.e. the scattering rate smaller than the spin-orbit energy~\cite{Michaeli2012-gl,Dimitrova2007-hp,Houzet2015-iy,Samokhin2008-nv,Smidman2017-hb}.
Thus, the cossover of the helical superconductivity and the sign reversal of $\Delta j_{\rm c}$ accompanied by it {may disappear} in the dirty-limit noncentrosymmetric superconductors, as shown near the transition temperature~\cite{Ilic2022-kh}. 
{A comprehensive study on the impurity effects on the SDE will be presented elsewhere~\cite{Ikeda_private}.}

We also comment on a technical aspect in determining the critical current.
In this paper, we considered superconducting solutions satisfying both $\Delta(q)\neq0$ and $F(q)<0$ to evaluate the critical current.
Strictly speaking, the condition $F(q)<0$ might be a kind of working hypothesis which reduces the numerical efforts.
The condition $F(q)<0$ corresponds to the smaller free energy in the superconducting state than in the zero-current normal state.
However, in principle, the comparison should be made with the current-flowing normal state, which is unfortunately out of equilibrium.
For this reason, it would be safe to understand the results for the depairing critical current as the limit of (meta)stability when the current is increased from inside the superconducting state.
When considering in this way, there seems to be no reason to discard the states with $F(q)>0$ to evaluate the critical current.
Note that the presence or absence of the condition $F(q)<0$ does not affect the results under low and moderate magnetic fields (except for the region where both the temperature and magnetic field are tiny), 
while
quasi-stable superconducting solutions with $F(q)>0$ actually exist under high magnetic fields.
Inclusion of these solutions will make small quantitative changes in the results, but we have confirmed that the qualitative results such as sign reversals of the SDE and re-entrant transition lines, are unchanged.

Finally, we comment on the nonreciprocal transition lines.
In the previous section, we have established the nonreciprocal transition lines in the presence of the supercurrent as a complementary probe for the SDE.
It should be noted that the phenomenology discussed in Sec.~\ref{subsec:phenomenology_critical_field} remains valid for the mechanisms of the SDE other than the intrinsic SDE as well, as long as the definitions Eqs.~\eqref{eq:Hc2p} and \eqref{eq:Hc2m} are valid.
To distinguish the intrinsic SDE from the other possible mechanisms, qualitative features such as the temperature scaling $r(T,h)\propto \sqrt{T_{\rm c}-T}$ near the transition temperature play an important role. 
It is also an interesting future issue to study the nonreciprocal transition lines caused by the extrinsic mechanisms of SDE.

\section{Summary}
\label{sec:summary}
In this paper, we have studied the superconducting diode effect and nonreciprocal phase transitions in noncentrosymmetric superconductors.
{Generalizing our previous paper,} we have derived a GL formula for the intrinsic SDE under low magnetic fields, which is applicable to {arbitrary}
noncentrosymmetric point groups.
The coupling of the effective spin-orbit coupling {of Cooper pairs} with the magnetic field determines the SDE, giving a convenient criterion to obtain a finite SDE for the given magnetic-field, crystal-axis, and electric-current directions.
The SDE of the Rashba-Zeeman model is also discussed for the $s$-wave and $d$-wave pairing symmetries.
The SDE in the $d$-wave superconducting state shares the qualitative features with the $s$-wave superconducting state, and show sign reversals as increasing the magnetic field.
The onset of the sign-reversed region almost coincides with the crossover line of the helical superconductivity, establishing the SDE as the promising probe of the helical superconductivity regardless of the pairing symmetry.
{Interestingly,} a larger nonreciprocity tends to be obtained in the $d$-wave state than in the $s$-wave state.

We have also {studied}
the nonreciprocity of the transition lines in the temperature-magnetic-field phase diagram, 
{pointing out their}
different behavior under the positive and negative current directions.
This can be observed either by directly determining the transition lines under a finite supercurrent or by drawing the contour plot of the critical current $j_{{\rm c} \pm}(T,H)$.
We have established the phenomenology of the nonreciprocal transition lines, which remains valid regardless of the microscopic origins of the nonreciprocity: There is a one-to-one correspondence between the SDE and the nonreciprocal transition lines when the transition lines are located where the SDE is small.
The skewness and crossings of the transition lines appear as typical behaviors.
The phenomenological results are illustrated {in the small- and large-current regions with the Rashba-Zeeman models for the $s$-wave and $d$-wave superconducting states.}
In contrast to the
{weakly nonreciprocal} region, the correspondence between the SDE and nonreciprocal transition lines might break down where the SDE is large.
There appears a kink in the transition lines under moderate electric current in Rashba-Zeeman superconductors, {and the transition lines can be even re-entrant.}
{We have also shown the first order transition and crossover in the superconducting state under the supercurrent.}

The sign reversals of the SDE as well as the anomalous behaviors of the nonreciprocal transition lines appear in the crossover region of helical superconductivity, where the Cooper-pair-momentum drastically changes. The re-entrant superconducting transition, first-order transition, and crossover occur in this region under the supercurrent.
Observations of these characteristics will explore the supercurrent-induced phenomena in superconductors and will provide an essential experimental hint to clarify the Cooper-pair-momentum related properties of helical superconductors.

\begin{acknowledgments}
We thank fruitful discussion with Teruo Ono, Yuta Miyasaka, Ryo Kawarazaki, Hideki Narita, Yuhei Ikeda, and Jun Ishizuka.
We also thank Hikaru Watanabe for helpful discussion and informing us of Ref.~\onlinecite{Levitov1985-pm}.
This work was supported by JSPS KAKENHI (Grants Nos. JP18H01178, JP18H05227, JP19H05825, JP20H05159, JP21K13880, JP21K18145, JP22H01181, JP22H04933) and SPIRITS 2020 of Kyoto University.
\end{acknowledgments}

\appendix
\section{Derivation of Eq.~\eqref{eq:Djc_GL}}
\label{app:GL_derivation}
Here we show the derivation of Eq.~\eqref{eq:Djc_GL} following Ref.~\onlinecite{Daido2022-ox}.
We write $\epsilon_i=1/2m_i$ for simplicity, and rescale the momentum as $Q_i=\sqrt{\epsilon_i}\delta q_i$.
In the absence of $\bm{g}_1$ and $\bm{g}_3$, the GL coefficients read
\begin{equation}
    \alpha=\alpha_0+\bm{Q}^2,\quad \beta=\beta_0.
\end{equation}
The electric current along the unit vector $\hat{n}$ is given by
\begin{equation}
    j_n(Q)=\hat{n}\cdot2\partial_{\bm{q}}f(\bm{q})=2\bm{n}'\cdot\partial_{\bm{Q}}f.
\end{equation}
Here, we defined $n'_i=\sqrt{\epsilon_i}\hat{n}_i$.
The corresponding unit vector is $\hat{n}'=\bm{n}'/\sqrt{\epsilon(\hat{n})}$, with $\epsilon(\hat{n})\equiv\sum_i\epsilon_i\hat{n}_i^2$.
Following the derivation for the conventional superconductors~\cite{Tinkham2004-dh,Daido2022-ox},
the critical current is achieved at the critical momenta $\bm{Q}_c=Q_{\rm c}\hat{n}'$ and $-\bm{Q}_c$, with
\begin{equation}
    Q_{\rm c}\equiv\sqrt{\frac{-\alpha_0}{3}}.
\end{equation}
Accordingly, we obtain
\begin{align}
    \bar{j}_{\rm c}(\hat{n})=j_n(Q_{\rm c})&=\frac{8\sqrt{\epsilon}}{3\sqrt{3}\beta_0}(-\alpha_0)^{3/2}.
\end{align}

Let us consider the first-order change of the critical current by the inclusion of $\bm{g}_1\cdot\bm{h}$ and $\bm{g}_3\cdot\bm{h}$. They cause the change of the dispersion $j_n(\bm{Q})\to j_n(\bm{Q})+\delta j_n(\bm{Q})$, and thus we have to evaluate $\delta j_n(\bm{Q}_c)$.
On the other hand, the change of the critical momentum $Q_{\rm c}$ does not contribute up to first order in $h$, owing to the definition of the critical current, $\partial_{\bm{Q}}j_n(\bm{Q})=0.$~\cite{Daido2022-ox}
By using
\begin{align}
    j_{{\rm c} +}&=j_n(\bm{Q}_{\rm c})+\delta j_n(\bm{Q}_{\rm c}),\\
    j_{{\rm c} -}&=j_n(-\bm{Q}_{\rm c})+\delta j_n(-\bm{Q}_{\rm c})\notag\\
    &=-j_n(\bm{Q}_{\rm c})+\delta j_n(\bm{Q}_{\rm c}),
\end{align}
($\delta j_n(\bm{Q})$ is an even function of $\bm{Q}$, as we see below) the nonreciprocity of the critical current is obtained by
\begin{align}
\Delta j_{\rm c}(\hat{n})=2\delta j_n(\bm{Q}_{\rm c}).
\end{align}

The expression of $\delta j_n(\bm{Q})$ is obtaind as follows.
By optimizing the order parameter, the free energy becomes
\begin{align}
    2\beta_0f(\bm{q})&=-2\beta_0\frac{\alpha(\bm{q})^2}{2\beta(\bm{q})}\notag\\
    &=-A^2(1-\bm{g}_1\cdot\bm{h})-2A\bm{g}_3\cdot\bm{h}+O(h^2),
\end{align}
where $A\equiv \alpha_0+\bm{Q}^2$.
Thus, we obtain
\begin{align}
    2\beta_0\delta j_n(\bm{Q})&=2\bm{n}'\cdot\partial_{\bm{Q}}\Bigl[A^2\bm{g}_1\cdot\bm{h}-2A\bm{g}_3\cdot\bm{h}\Bigr].
\end{align}
Note that this is an even function of $\bm{Q}$.
By substituting $\bm{Q}_{\rm c}$, we obtain
\begin{align}
    \Delta j_{\rm c}(\hat{n})=\frac{8\alpha_0^2}{9\beta_0}\,\bm{g}_{\mathrm{eff}}(\hat{n})\cdot\bm{h},
\end{align}
after some algebra. {Here we used} the relations such as $\delta q_i(\bm{Q})\equiv Q_i/\sqrt{\epsilon}_i$ and
\begin{align}
    &\hat{n}'\cdot\partial_{\bm{Q}_{\rm c}}\bm{g}_3(\delta\bm{q}(\bm{Q}_{\rm c}))\notag\\
    &=\lim_{\eta\to0}\frac{\bm{g}_3(\delta\bm{q}(Q_{\rm c}\hat{n}'+\eta Q_{\rm c}\hat{n}'))-\bm{g}_3(\delta\bm{q}(Q_{\rm c}\hat{n}'))}{Q_{\rm c}\eta}\notag\\
    &=\lim_{\eta\to0}\frac{Q_{\rm c}^3(1+\eta)^3\bm{g}_3(\delta\bm{q}(\hat{n}'))-Q_{\rm c}^3\bm{g}_3(\delta\bm{q}(\hat{n}'))}{Q_{\rm c}\eta}\notag\\
    &=3Q_{\rm c}^2\bm{g}_3(\hat{n})/\sqrt{\epsilon(\hat{n})^3},
\end{align}
where $\delta\bm{q}(\hat{n'})=\hat{n}/\sqrt{\epsilon(\hat{n})}$ and $\bm{g}_3(a\bm{q})=a^3\bm{g}_3(\bm{q})$. 
Thus, we obtain
\begin{align}
    r&=\frac{1}{2\sqrt{3}}\sqrt{\frac{-\alpha_0}{\epsilon(\hat{n})}}\bm{g}_{\mathrm{eff}}(\hat{n})\cdot\bm{h}.
\end{align}

\section{GL theory of the SDE {beyond the limit of small} magnetic fields}
\label{app:GL_general_h}

\begin{figure}[t]
    \centering
    \includegraphics[width=0.45\textwidth]{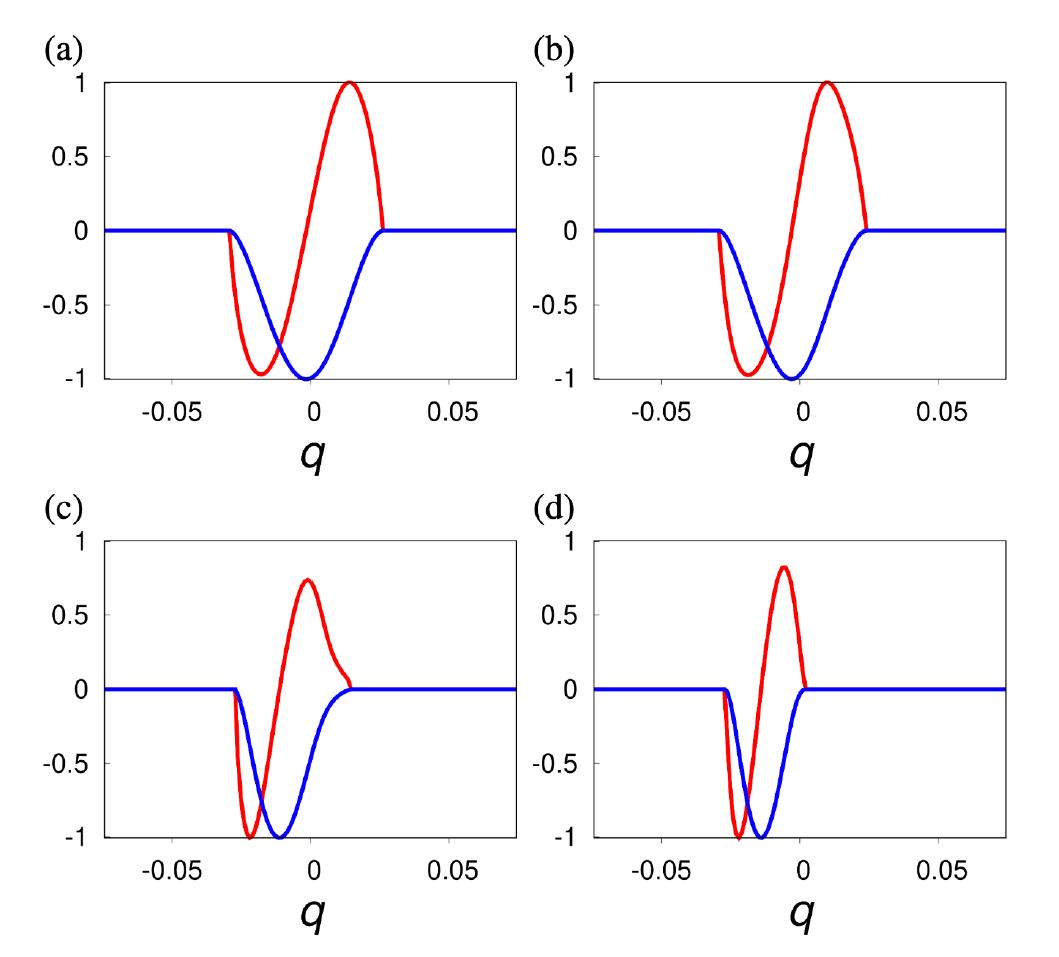}
    \caption{Momentum dependence of the supercurrent $j(q)$ [red color] and the condensation energy $F(q)$ [blue color] for the $s$-wave state at $T=0.02$.
    $j(q)$ and $F(q)$ are normalized by the maximum of their absolute value.
Calculations are done with 
    (a) $h=0.04$, (b) $h=0.05$, (c) $h=0.0585$, and (d) $h=0.06$.
    }
    \label{fig_app:Fq_jq_s}
\end{figure}
\begin{figure}[t]
    \centering
    \includegraphics[width=0.45\textwidth]{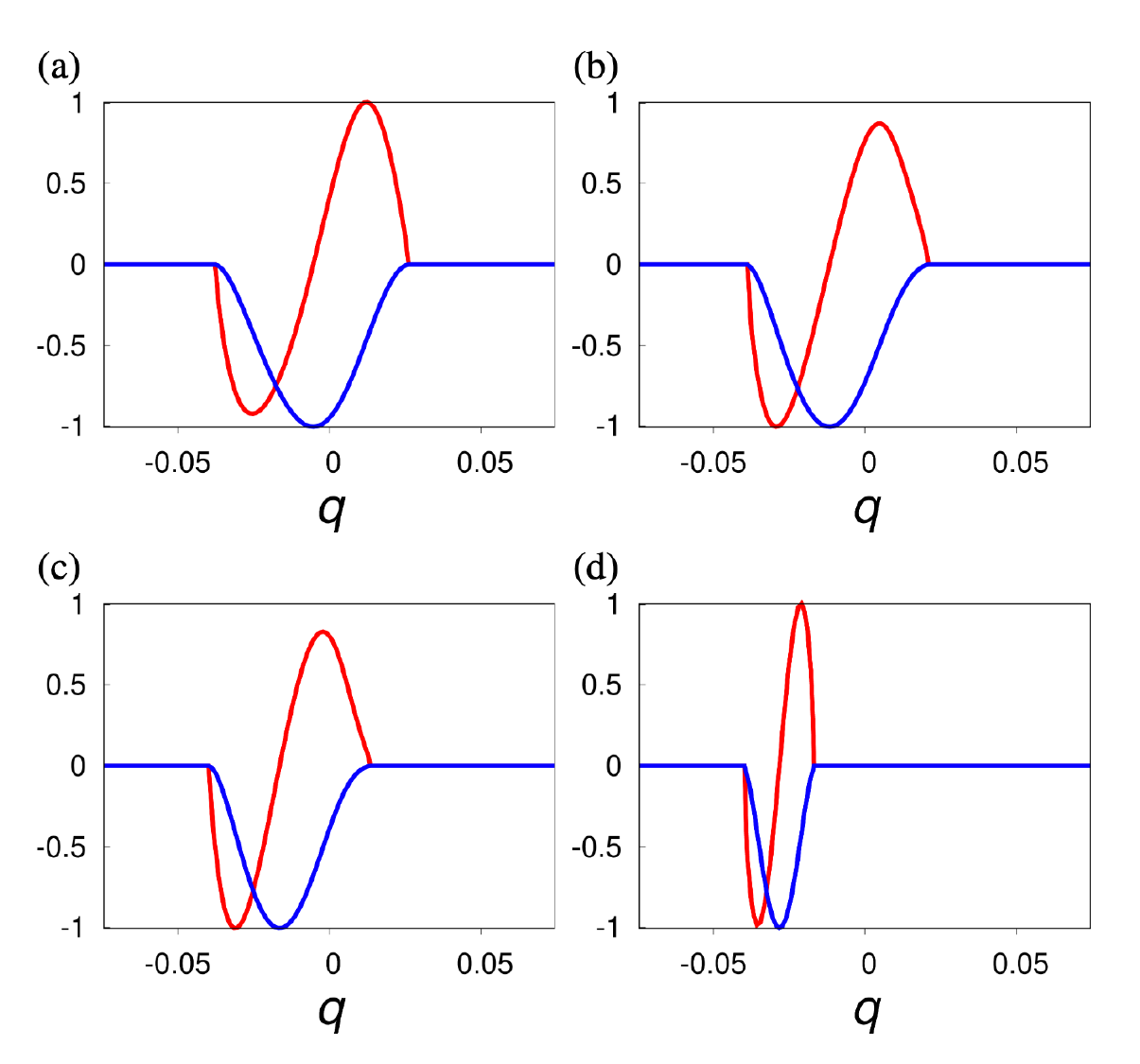}
    \caption{$j(q)$ and $F(q)$ for the $d$-wave state at $T=0.02$.
    Notations are the same as those of Fig.~\ref{fig_app:Fq_jq_s}.
    Calculations are done with (a) $h=0.04$, (b) $h=0.05$, (c) $h=0.055$, and (d) $h=0.07$.}
    \label{fig_app:Fq_jq_d}
\end{figure}

Here we show that the temperature scaling of the SDE $\Delta j_{\rm c}\propto(T_{\rm c}-T)^2$ and $Q\propto\sqrt{T_{\rm c}-T}$ holds even when higher-order effects of the magnetic field is considered.
Let us start from {the GL free energy}
\begin{align}
    f(q,\psi)&=\alpha(q)\psi^2+\frac{1}{2}\beta(q)\psi^4.
\end{align}
For simplicity, we consider a scalar {momentum} $q$ instead of the vector one.
The order parameter has a nontrivial solution when
\begin{align}
    \psi^2=-\alpha(q)/\beta(q)>0.
    \label{eq:appendixB2}
\end{align}
By assuming the second-order phase transition and thus $\beta(q)>0$ for all the $q$ values we are interested in, the superconducting transition is triggered by the sign reversal of $\alpha(q)$ for some $q$.
This means that the minimum of $\alpha(q)$ changes from positive to negative {at the transition temperature $T_{\rm c}$,} leading to
\begin{align}
\alpha(q)=-\tau+\alpha_2(q-q_\alpha)^2+\alpha_3(q-q_\alpha)^3+\cdots,   
\end{align}
and $\tau\propto T_{\rm c}-T$ is positive {when Eq.~\eqref{eq:appendixB2} is satisfied.}
By setting $\delta q\equiv q-q_\alpha$ and expanding $\beta(q)$ by $\delta q$, we obtain
\begin{align}
f(q)&=-\frac{(-\tau+\alpha_2\delta q^2+\alpha_3\delta q^3+\cdots)^2}{2\beta_0(1+\beta_1\delta q+\cdots)}\label{eq:f_temp_1},\\
&=-\frac{\tau^2}{2\beta_0}\frac{(1-\alpha_2\delta q_\tau^2-\sqrt{\tau}\alpha_3\delta q^3_\tau+O(\tau))^2}{1+\sqrt{\tau}\beta_1\delta q_\tau+O(\tau)},
\end{align}
by introducing a new variable $\delta q_\tau=\delta q/\sqrt{\tau}$.
Note that $f(q)$ vanishes for $\tau\to0$ at $\delta q_\tau=1/\sqrt{\alpha_2}=O(1)$.
Thus, we are interested in the region where $\delta q_\tau=O(1)$, and the higher order terms in the expansion of $\delta q$ are negligible compared to the $\alpha_3$ and $\beta_1$ terms.
Since 
they are multiplied by the small quantity $\sqrt{\tau}$, it is sufficient to consider their first-order perturbation to discuss the temperature scaling of the SDE.
Equation~\eqref{eq:f_temp_1} formally coincides with the GL free energy considered for the $O(h)$ SDE in the main text and Ref.~\onlinecite{Daido2022-ox}, and therefore we obtain
\begin{align}
    \Delta j_{\rm c}\propto \tau^2,\quad Q\propto \sqrt{\tau},
\end{align}
for this case as well.
We also obtain $q_0=q_\alpha+O(\tau)$ for the helical superconductivity, since $f(q)$ is minimized at $\delta q_\tau=O(\sqrt{\tau})$.

\section{Momentum dependence of the condensation energy at a high temperature}
\label{app:fig_high_T}

We show in Figs.~\ref{fig_app:Fq_jq_s} and \ref{fig_app:Fq_jq_d} the results of $F(q)$ {and $j(q)$} at $T=0.02$ for the $s$-wave and $d$-wave states under various magnetic fields.
The condensation energy $F(q)$ shows a single-well structure, in contrast to that at low temperatures.


%


\end{document}